\documentclass[11pt]{article}
\usepackage[margin=1in]{geometry}
\usepackage{amsmath,amssymb,amsthm,mathtools,booktabs}
\usepackage[T1]{fontenc}
\usepackage{lmodern,microtype}
\usepackage{caption}
\usepackage[hidelinks]{hyperref}
\usepackage{aliascnt}
\usepackage{cleveref}
\let\cref\Cref
\hypersetup{pdftitle={Affine Combiners for Reed--Solomon Proximity Gaps},pdfauthor={Alessandro Chiesa; Guy Weissenberg; Eylon Yogev; Kai Zhe Zheng},pdfsubject={Finite affine MCA bounds and a separation from scalar combiners},bookmarksopen=true}
\newtheorem{theorem}{Theorem}[section]
\newcommand{\newsharedtheorem}[2]{%
  \newaliascnt{#1}{theorem}\newtheorem{#1}[#1]{#2}\aliascntresetthe{#1}}
\newsharedtheorem{lemma}{Lemma}
\newsharedtheorem{claim}{Claim}
\newsharedtheorem{corollary}{Corollary}
\newsharedtheorem{proposition}{Proposition}
\theoremstyle{remark}\newsharedtheorem{remark}{Remark}
\crefname{claim}{Claim}{Claims}
\newcommand{\F}{\mathbb F}
\newcommand{\Err}{\varepsilon}
\newcommand{\calD}{\mathcal D}
\newcommand{\ceil}[1]{\left\lceil #1\right\rceil}
\newcommand{\floor}[1]{\left\lfloor #1\right\rfloor}
\newcommand{\Res}{\operatorname{Res}}
\newcommand{\mc}{\mathcal}
\newcommand{\eps}{\varepsilon}
\newif\ifrmext
\newcommand{\rmon}{\rmexttrue}

\rmon
\newcommand{\FormatAuthor}[3]{%
  \begin{tabular}[t]{@{}c@{}}
    #1\\[0.2em]
    {\small #3}\\
    {\small\href{mailto:#2}{\texttt{#2}}}
  \end{tabular}%
}
\title{Affine Combiners for Reed--Solomon Proximity Gaps}
\author{%
  \begin{tabular}{@{}c@{\hspace{3em}}c@{}}
    \FormatAuthor{Alessandro Chiesa}{alessandro.chiesa@epfl.ch}{EPFL} &
    \FormatAuthor{Guy Weissenberg}{guy.weissenberg@epfl.ch}{EPFL}
    \\\noalign{\vskip 1em}
    \FormatAuthor{Eylon Yogev}{eylon.yogev@biu.ac.il}{Bar-Ilan University} &
    \FormatAuthor{Kai Zhe Zheng}{kzzheng@mit.edu}{Simons Institute for the Theory of Computing}
  \end{tabular}%
}

\theoremstyle{definition}
\newsharedtheorem{definition}{Definition}
\numberwithin{equation}{section}
\AddToHook{cmd/section/before}{\clearpage}
\AddToHook{env/thebibliography/begin}{\phantomsection\addcontentsline{toc}{section}{\refname}}
\begin{document}
\pagenumbering{roman}
\pagestyle{empty}
\maketitle
\thispagestyle{empty}
\vspace{-2em}
\begin{abstract}
Reed--Solomon proximity gaps establish soundness guarantees for protocols that
test received words through random combinations. For length-$n$ codes
over $\F_q$, classical scalar combiners achieve error $O(n/q)$ in the
unique-decoding regime, with larger known bounds near the Johnson radius.

We prove smaller error bounds using affine combiners, which combine
two received words $f,g$ as
$f(x)+(\alpha+\beta x)g(x)$, where $\alpha,\beta$ are independent
uniform elements of $\F_q$. For output degree bound $d=\rho n$,
fixed $\rho\in(0,1)$ and $0<\eta<1-\sqrt\rho$, and sufficiently
large block length $n$, we prove mutual correlated agreement (MCA)
with one coordinate of agreement loss at threshold
$A=\ceil{(\sqrt\rho+\eta)n}$, with error
\[
O_{\rho,\eta}\!\left(\frac1q+\frac{n^2}{q^2}\right).
\]
The result holds over arbitrary evaluation domains in every finite
field, including prime and binary extension fields. The agreement
threshold can approach the Johnson threshold, beyond unique decoding.

For $q\ge n^2$, our error is $O_{\rho,\eta}(1/q)$, an asymptotic
factor-$n$ separation from scalar combiners: we prove that their
worst-case MCA error is $\Omega_{\rho,\eta}(n/q)$ at the same agreement
threshold and with the same one-coordinate loss. We also prove that
affine combiners incur worst-case error $\Omega_{\rho,\eta}(n/q)$
when no agreement loss is allowed. Thus one coordinate of agreement
loss is necessary for our guarantee; this relaxation of standard MCA
reduces the guaranteed common agreement fraction by at most $1/n$.
We also extend the result to $k$-variate Reed--Muller codes on subfield
grids, obtaining MCA error $O_{\rho,\eta,k}(q^{-k})$ for fixed
$\rho,\eta,k$, sufficiently large $n$, and $q\ge n^{k+1}$, with
agreement loss $n^{1-1/k}$.


\end{abstract}

\clearpage
\tableofcontents

\clearpage
\pagenumbering{arabic}
\pagestyle{plain}
\section{Introduction}
Reed--Solomon proximity testing is a central component of efficient
proof systems, including FRI~\cite{FRI}, STIR~\cite{STIR}, and WHIR~\cite{WHIR}.
Many protocols test a batch of received words by checking a random linear
combination of them for proximity to a Reed--Solomon code. A word is
close to the code when it agrees with a low-degree polynomial on many
evaluation points. The soundness of this approach rests on a
\emph{proximity gap} guarantee which is typically a statement of the following form: if at least one of the words in the combination is $\delta$-far from the code, then the combination is still $\delta'$-far from the code except with a small error probability $\varepsilon$. Typically one wants $\delta'$ either equal or at least close to $\delta$ and $\eps$ as small as possible.

Classical results establish such gaps for $\delta$ all the way up to the Johnson
radius~\cite{BCIKS,BCHKS,DKT}, but the error, $\eps$, that they achieve can sometimes be an issue. Indeed, for a
length-$n$ Reed--Solomon code over a field $\F_q$, the classical
proximity-gap bounds are $n/q$, even in the unique-decoding
regime~\cite[Theorem~1.2]{BCIKS}, and this dependence on the block length
costs $\log_2 n$ bits of soundness when used in proof systems. For many practical use cases, this can cost $20-30$ bits of soundness, forcing the proof system to compensate for this gap via other means such as working over a larger field or using more bits of grinding -- both of which can lead to a less efficient prover.

In this paper, we propose a simple variant of proximity gaps which uses \emph{affine combiners} and show that it can achieve error on the order of $1/q$, surpassing the $\Omega(n/q)$-error lower bound for the standard scalar combiners considered in classical proximity gaps. We describe our variant and main results in detail next.

\subsection{Our results}\label{sec:our-results}
Throughout this section fix an integer degree bound $1\le d<n$, a domain
$\calD\subseteq\F_q$ of size $n$, and received words
$f,g:\calD\to\F_q$. Set $\rho\coloneq d/n$ to be the rate.
Our affine combiners for proximity gaps work as follows. Randomly sample independent uniform
$\alpha,\beta\in\F_q$ and form the combined word
\[
 w_{\alpha,\beta}(x)\coloneq f(x)+(\alpha+\beta x)g(x),\qquad x\in\calD.
\]
We call the coefficient $\alpha+\beta x$ an \emph{affine combiner}.
Unlike scalar mixing, multiplication by this coefficient can increase
polynomial degree by one. To keep the combined polynomial's degree at
most $d$, we therefore use different degree bounds for the two words:
in the honest case, $f$ is the evaluation of a polynomial of degree at
most $d$, and $g$ of one of degree at most $d-1$.

For many proof systems, we actually require a stronger notion of proximity gaps called \emph{Mutual Correlated Agreement} (MCA) \cite{WHIR}. At a high level, mutual correlated agreement says that, except with a small error probability (called MCA error), the combined word does not have \emph{any} agreement domain $I$ (with a degree $d$ polynomial) which is not also a shared agreement domain of $f$ (with degree $d$) and $g$ (with degree $d-1$). The asymmetry in our degree bounds also leads to a modified definition of MCA error which we define as follows.

\paragraph{Definition of the MCA error for affine combiners.}
Let $1\le A\le n$ be an integer agreement threshold, and let the
agreement loss $s\ge0$ be an integer. Define
\[
 S_{\alpha,\beta}(P)
 \coloneq\{x\in\calD:P(x)=f(x)+(\alpha+\beta x)g(x)\}.
\]
For $(\alpha,\beta)$, polynomials $F,G\in\F_q[X]$ of degrees at most $d$ and $d-1$, respectively,
are admissible witnesses for $P=F+(\alpha+\beta X)G$ with agreement loss at most $s$ if
\[
\begin{gathered}
 \bigl|\{x\in S_{\alpha,\beta}(P):
 F(x)\ne f(x)\text{ or }G(x)\ne g(x)\}\bigr|\le s.
\end{gathered}
\]
We say that the affine combiner has MCA error at most $\varepsilon$ with
agreement loss at most $s$ if, for every pair $f,g$,
\[
 \Pr_{\alpha,\beta\,\leftarrow\,\F_q}
 \left[\begin{array}{l}
 \exists P\in\F_q[X],\ \deg P\le d:\\
 |S_{\alpha,\beta}(P)|\ge A\text{ and }P\text{ has no}\\
 \text{admissible witnesses with agreement loss at most }s
 \end{array}\right]\le\varepsilon.
\]
We denote the smallest such $\varepsilon$ by
$\Err_{\rm aff}(n,\rho,A,s,q)$. Our main theorem shows that in the agreement loss $s = 1$ case, MCA error is $O_{\rho,\eta}(1/q+n^2/q^2)$. In this case we abbreviate the error as $\Err_{\rm aff}$. For fixed $\rho,\eta$ and $q\ge n^2$, the error is
$O_{\rho,\eta}(1/q)$, removing the factor $n$ which is necessary for scalar proximity gaps (see \cref{sec:separation}). Our results hold over all fields.

\begin{theorem}\label{thm:main}
Fix $\rho\in(0,1)$ with $\rho n$ an integer, let
$0<\eta<1-\sqrt\rho$, and set $A\coloneq\ceil{(\sqrt\rho+\eta)n}$.
Suppose $n\eta^2\ge4/\rho$. Then, for every finite field $\F_q$ with
$q\ge n$ and every fixed domain $\calD\subseteq\F_q$ of size $n$,
the MCA error with agreement loss $s=1$ satisfies
\begin{equation}\label{eq:main-asymptotic}
 \Err_{\rm aff}\Bigl(
 \underbrace{n}_{\substack{\text{block}\\\text{length}}},
 \underbrace{\rho}_{\substack{\text{degree}\\\text{fraction}}},
 \underbrace{A}_{\substack{\text{agreement}\\\text{threshold}}},
 \underbrace{1}_{\substack{\text{agreement}\\\text{loss}}},
 \underbrace{q}_{\substack{\text{field}\\\text{size}}}
 \Bigr)
 \le\frac{4}{\eta q}+\frac{2^{14}n^2}{\rho^3\eta^{10}q^2}.
\end{equation}
\end{theorem}

\paragraph{Small error beyond unique decoding.}
Taking the 128-bit binary field $\F_{2^{128}}$, block length $n=2^{20}$,
rate $\rho=1/4$, and agreement margin $\eta=0.05$ (so $A=576717$), the theorem gives
\[
 \Err_{\rm aff}(2^{20},1/4,576717,1,2^{128})
 \le \frac{80}{2^{128}}+\frac{5^{10}}{2^{176}}
 \approx 2^{-122}.
\]
Thus, with agreement loss $s=1$, the MCA error bound is approximately
$2^{-122}$. In contrast, the classical scalar proximity-gap bound is
$n/q=2^{-108}$ even in the unique-decoding
regime~\cite[Theorem~1.2]{BCIKS}.

\paragraph{Dealing with agreement loss in IOPPs.}
Moving from $s=0$ to $s=1$ reduces the guaranteed common agreement
fraction by at most $1/n$ after the combination. Protocols can account
for the agreement loss by making more queries. For example, suppose that, in a soundness
argument, we bound the probability that $t$ independent uniform queries
all land in an agreement set by $a^t$, where $0<a\le1-1/n$.
If the only change is agreement loss $s=1$, the corresponding
bound becomes $(a+1/n)^t$, with multiplicative increase at most
\[
 \frac{(a+1/n)^t}{a^t}
 =\left(1+\frac1{an}\right)^t
 \le \exp\!\left(\frac{t}{an}\right).
\]
For a constant agreement fraction and a query count much smaller than
$n$, this has a negligible effect on the sampling contribution to
soundness. For $n=2^{20}$, $t=128$, and $a\ge1/2$, the soundness cost is less
than $0.00036$ bits. 

\paragraph{Equal degree bound.}
For applications requiring equal input degree bounds, we give a simple
modification. Sample an additional
independent uniform $\theta\in\F_q$ and instead form
\[
 w_{\theta,\alpha,\beta}(x)
 \coloneq g(x)+(\alpha+\beta x)(f(x)+\theta g(x)).
\]
Both witness polynomials may now have degree at most $d-1$, while
the combined polynomial still has degree at most $d$.
In Section~\ref{sec:equal-degree}, we show that this modification
achieves essentially the same MCA error.

\paragraph{Application: smaller fields for Flock and leanVM.}
Flock~\cite{Flock} uses the Ligerito polynomial commitment scheme over
$\F_{2^{128}}$, where the block-length factor, $n$, in scalar MCA bounds
creates a substantial soundness gap at the $128$-bit security target.
Moving to a larger field increases arithmetic and communication costs;
leanVM uses a $192$-bit binary field~\cite{leanVM}.
Affine combiners remove this factor $n$ from the MCA error and could support
efficient $128$-bit arithmetic in these systems. For the representative
parameters above, Remark~\ref{rem:flock-folding} bounds the per-round MCA
contribution below $2^{-128}$ with $3$--$5$ grinding bits per fold challenge,
depending on the variant. Establishing $128$-bit security for a full system
additionally requires accounting for composition across folds, query errors,
and the security of the commitments and Fiat--Shamir transformation.
Section~\ref{sec:equal-degree} develops the adaptations for linear coefficient
claims, degree growth, and shared agreement across interleaved rows needed
to integrate affine combiners into these proof systems.

\paragraph{Example: the toy protocol.}
To further demonstrate the benefits of affine combiners, we consider the toy protocol
in~\cite[Construction~6.2]{ABF26} with evaluation at zero as the linear constraint.
The verifier receives $u,v\in\F_q$ and has oracle access to received
words $f,g:\calD\to\F_q$, where $|\calD|=n$ and $0\notin\calD$.
The honest prover knows polynomials $F,G\in\F_q[X]$
satisfying $F(0)=u$, $G(0)=v$, and
\[
 \deg F\le d,\quad \deg G\le d-1,\qquad
 F|_{\calD}=f,\quad G|_{\calD}=g.
\]
The verifier samples independent uniform $\alpha,\beta\in\F_q$,
and the prover sends the coefficients of a candidate polynomial $P$
with $\deg P\le d$. The verifier then samples $t$ independent uniform
coordinates $x_1,\ldots,x_t\in\calD$, with replacement, and accepts if
\[
 P(0)=u+\alpha v,\qquad
 P(x_i)=w_{\alpha,\beta}(x_i)\quad\text{for every }i.
\]
For the honest response $P=F+(\alpha+\beta X)G$, acceptance is certain.

We denote by $\ell$ the number of polynomial pairs with the stated degree
bounds that agree with $f,g$ on at least $A-1$ common coordinates.
If no such pair satisfies both evaluation claims, then by
Theorem~\ref{thm:main},
\begin{equation}\label{eq:toy-affine-soundness}
 \Pr[\text{accept}]
 \le \Err_{\rm aff}(n,\rho,A,1,q)+\frac{\ell}{q}
      +\left(\frac{A-1}{n}\right)^t.
\end{equation}
Indeed, for each pair that does not satisfy both evaluation claims,
$(F(0)-u)+\alpha(G(0)-v)=0$ for at most $q$ choices of $(\alpha,\beta)$.
After excluding at most $\ell q$ such challenges and the MCA exceptional set,
$|S_{\alpha,\beta}(P)|\le A-1$ whenever $P(0)=u+\alpha v$.
By independence of the queries, we obtain~\eqref{eq:toy-affine-soundness},
as in~\cite[Lemmas~6.6 and~6.8]{ABF26}.

Concretely, for $n=2^{14}$, $d=n/4-1$, $A=\ceil{0.55n}$, and $t=160$,
$\eta=0.55-\sqrt\rho>0.05$ and $\ell\le5$ by the Johnson bound.
For scalar combiners, with
degree bounds $\deg F,\deg G\le d$, agreement loss $s=0$, and the
same $n,A,t$, we obtain the soundness error bound
$2^{30}/q+0.55^{160}$ from~\cite[Theorem~4.12]{ABF26}.
Without grinding, the soundness error is less than $2^{-128}$ over
$\F_{2^{136}}$ with affine combiners and over $\F_{2^{160}}$ with scalar combiners.
The corresponding response sizes are $68$ KiB and $80$ KiB,
a $15\%$ reduction with affine combiners.
For affine combiners, the verifier sends one additional field element and
performs one additional multiplication per queried coordinate.
We require $\deg G\le d-1$ and allow agreement loss $s=1$, keeping
the output degree bound $d$ and query count $t$ in both protocols.

\ifrmext
\paragraph{Extension to Reed--Muller codes.}
For Reed--Muller codes, which consist of evaluations of multivariate
low-degree polynomials, we obtain MCA error \emph{even below $1/q$}.
Their multivariate structure gives us more degrees of freedom in the
random combination. For $k$-variate codes, we use an affine combiner
$\alpha_0+\alpha_1x_1+\cdots+\alpha_kx_k$, with independent uniform
coefficients in $\F_q$.
For fixed degree fraction $\rho$, Johnson slack $\eta$, and number of
variables $k$, Section~\ref{sec:rm} gives error
$O_{\rho,\eta,k}(q^{-k})$ for sufficiently large block length $n$ and
$q\ge n^{k+1}$. We use the agreement threshold of
Theorem~\ref{thm:main} and allow agreement loss $n^{1-1/k}$.
Thus the number of variables becomes the error exponent, giving
$o(1/q)$ error for every fixed $k\ge2$. We believe that this phenomenon is related to the rich invariance structure of Reed--Muller codes and it would be interesting to see if proximity gaps (via some modified combination like here) below $1/q$ could be shown for any other codes with a large group of invariances.
\fi


\subsection{Related Work and Further work}
We discuss a few lines of work related to the questions we consider.
\paragraph{Progress and limits beyond Johnson.}
Recent work substantially extends the range of agreement thresholds
in large characteristic. Building on the list-decoding breakthrough
of Brakensiek, Chen, Putterman, Zhang, and Zheng~\cite{BCPZZ},
Jeronimo~\cite{Jer26} proves exact-support MCA over prime fields on
arbitrary evaluation domains, up to any fixed slack $\gamma>0$ from
capacity. For scalar lines, the error is $n^{O_\gamma(1)}/q$, with
agreement loss $s=0$. Dao, Kominers, and Thaler~\cite[Theorem~1.1]{DKT}
give a first-order agreement threshold $a_1(\rho)<\sqrt{\rho}$ in
sufficiently large characteristic. At agreement $a_1(\rho)+\eta_1$ with
$\eta_1>0$, they bound the list size by $O_\rho(n/\eta_1^2)$ and the
scalar MCA error by $O_\rho(n^2/(q\eta_1^4))$. They also give polynomial
list-size and exceptional-challenge bounds up to any fixed gap from
capacity~\cite{DKT}. Gabizon~\cite{Gab26} proves a
linear list-size bound at an agreement threshold strictly below Johnson
over fields of sufficiently large characteristic. Harsha, Kumar, and
Saptharishi~\cite{HKS26} give a unified exposition of the capacity
list-decoding and proximity-gap results.

Recent work establishes limits beyond the Johnson radius over fields
of small characteristic~\cite{DKTZ26,AGG26}. In particular, Dao, Kominers, Thaler,
and Zheng rule out a general $\operatorname{poly}(n)/q$ proximity-gap
error bound on additive binary domains. These results highlight
the distinction between extending the decoding radius and reducing
the error within a given radius. Our focus is on the latter: we
remain in the Johnson regime and prove a small, explicit error bound
over every finite field by using affine combiners with
agreement loss $s=1$.

We remark that we can also extend our analysis to achieve slightly better MCA error beyond Johnson using the one derivative interpolator, analogous to the (slightly) beyond Johnson proximity gap results in \cite{DKT,Gab26}, but leave it to future work to give a clean writeup.

\paragraph{Concurrent work.}
Frigo and shelat's \emph{Polygero}~\cite{FS26} consider a version of proximity gaps that uses polynomial coefficients. Instead of one scalar for each
received word $w_i$, the verifier samples an independent polynomial
$B_i$ of degree at most $r$ and tests the combination $\sum_i B_iw_i$. Roughly speaking, their analysis gives MCA error $O(1/q)$ with agreement
loss up to $r$ coordinates using degree $r=\Omega(n/\log_2 q)$ challenges~\cite[Section~7]{FS26}.
As a concrete example, \cite[Remark~2]{FS26} uses degree-$74$ challenges for a
code of length $25724$ and rate about $1/16$ over $\F_{2^{128}}$,
with $64$ queried coordinates and soundness error
at most $2^{-120}$. 

Our work shows that \emph{degree-one
challenges already give small error with agreement loss only one}.
At fixed rate and fixed slack from Johnson, we bound the probability
of an exceptional combiner by $O(1/q+n^2/q^2)$, and hence by $O(1/q)$
when $q\ge n^2$. 

\section{Techniques}\label{sec:techniques}
In this section, we overview our techniques and give some intuition on why affine combiners can
achieve error on the order of $O(1/q)$. We begin with the intuition, and our inspiration for considering affine combiners in the first place. For the remainder of the section, fix the evaluation domain $\calD$, a degree $d$, received words $f,g:\calD\to\F_q$ and set the agreement parameter to be $A=\ceil{(\sqrt\rho+\eta)n}$, where $\eta>0$ is a slack parameter and $\rho = d/n$ is the rate.

Our starting point is the following view of
proximity gaps as a low-degree test for a special type of input function. Specifically, from, $f,g$, form the table
\[
 H(x,z)\coloneq f(x)+zg(x),\qquad (x,z)\in\calD\times\F_q.
\]
In the honest case, when $f,g$ are evaluations of polynomials $F,G$
with $\deg F\le d$ and $\deg G\le d-1$, this table agrees with
$F(X)+ZG(X)$, a bivariate polynomial of total degree at most $d$
that is linear in $Z$. An important fact about low degree bivariate polynomials is that their restrictions to affine lines are low degree as well. Specifically, restricting $H$ to the affine line intersected with $\calD \times \F_q$,
\[
 \ell_{\alpha,\beta}\coloneq
 \{(x,\alpha+\beta x):x\in\calD\},
\]
gives the univariate function
\[
 x\longmapsto H(x,\alpha+\beta x)=f(x)+(\alpha+\beta x)g(x),
\]
which has degree at most $d$ in the honest case. Hence, these restrictions give exactly our affine combinations. On the other hand, scalar combinations correspond to restricting only to the axis-parallel lines $z=\alpha$.

Viewed this way, affine combiners are analogous to the affine line-versus-point
test of Arora and Sudan~\cite{AS03}, whereas scalar combiners are analogous to
the axis-parallel test of Polishchuk and Spielman~\cite{PS94}. The former admits analyses in the list-decoding
regime~\cite{HKSS24,KTZ}, whereas the classical Polishchuk--Spielman
analysis applies only in the unique-decoding regime. This gives some motivation for affine combiners potentially achieving smaller MCA error.

A natural question is then the following: if at least $\Omega(1/q)$ fraction
of the affine-line restrictions of $H$ above agree with degree-at-most-$d$ polynomials on
at least $A$ coordinates, does $H$ have comparably large agreement
with a global total-degree-$d$ polynomial that is linear in $Z$?
Such a polynomial would have the form $F(X)+ZG(X)$ with $F, G$ of degrees at most $d, d-1$ respectively, and it is not hard to see that concluding agreement of $H$ with such a polynomial would lead to correlated agreement of $f, g$ with $F, G$ respectively (up to a one coordinate loss).

To this end, Minzer and Zheng~\cite{MZ25}
give a motivating result for total-degree Reed--Muller codes. For an
absolute constant $\tau>0$, let $\varepsilon\ge(d/q)^\tau$ and
$u:\F_q^3\to\F_q$. Suppose that for more than a
$100/(\varepsilon^2q)$ fraction of affine planes $P\subseteq\F_q^3$,
the restriction $u|_P$ agrees with a polynomial of total degree at most
$d$ on at least a $1.2\varepsilon$ fraction of $P$. Then $u$ agrees
with a trivariate polynomial of total degree at most $d$ on more than
an $\varepsilon$ fraction of $\F_q^3$. Thus agreement on more than
$O_\varepsilon(1/q)$ of the plane restrictions forces global agreement, giving evidence that global structure can potentially be concluded from such a small fraction of restrictions. One can also check that a typical choice of $\Omega(1/q)$-fraction of affine lines intersected with $\calD \times \F_q$ covers a constant fraction of the domain $\calD \times \F_q$ whereas $\Omega(1/q)$-fraction of the axis parallel lines in $\calD \times \F_q$ covers only a negligible fraction of the domain.

\paragraph{Proof overview.}
The proof follows the same high-level strategy as in the scalar case, developed in \cite{BCIKS}, requiring mostly minor adaptations to handle the affine-combiners we consider. We heavily borrow tools from prior works \cite{BCIKS, BCHKS, DKT}.

We start by interpolating a low-degree ``explainer'' polynomial $Q(X,Y,Z)$ with the following property. For every $\alpha,\beta\in\F_q$ and polynomial
$P$ of degree at most $d$ whose agreement domain
$I=S_{\alpha,\beta}(P)$ has size at least $A$, we require
\[
 Q(X,\alpha+\beta X,P(X))=0.
\]
In particular, this identity holds for every \emph{exceptional triple}, i.e.\
a triple for which $I$ contains at least two points outside the
correlated agreement domain of every polynomial pair $F,G$ satisfying
$\deg F\le d$, $\deg G\le d-1$, and
$P=F+(\alpha+\beta X)G$.
We construct $Q$ by adapting the interpolation argument of \cite{BCHKS}. 

Once such a $Q$ is found, we count the pairs $(\alpha,\beta)$ for which this
identity has a polynomial solution $P$ forming an exceptional triple.
This counting is again done via standard techniques, this time via adaptations of the factorization and root lifting arguments from~\cite[Section~5]{BCIKS}
and the agreement-incidence counting technique from~\cite{DKT}.
For simplicity, assume that $Q$ is irreducible; otherwise we factor $Q$
and apply the argument to each irreducible factor. For factors of
positive $Z$-degree, we also describe the case $Q_Z\not\equiv0$, where
$Q_Z=\partial Q/\partial Z$. The proof handles factors with identically
zero derivative in small characteristic by a change of variables.
We partition the exceptional triples into three classes.
\begin{itemize}
 \item \emph{Triples over excluded coefficient pairs.}
 In this step, we first count the coefficient pairs $(\alpha,\beta)$
 occurring in roots $(\alpha,\beta,P)$ such that either
 \[
 \begin{aligned}
  Q(X,\alpha+\beta X,Z)&\equiv0,\qquad\text{or}\\
  Q_Z(X,\alpha+\beta X,P(X))&\equiv0.
 \end{aligned}
 \]
 The reason for filtering out these roots first is that our later
 arguments need both expressions above to be nonzero. If the first
 is zero, then the equation above holds for every $P$ and gives no information
 on which polynomial $P$ can be. For irreducible $Q$, this can happen
 only if $Q$ is independent of $Z$. Each such pair gives a linear
 factor $Y-\alpha-\beta X$ of $Q$. Distinct pairs give distinct factors
 of $Y$-degree one, so there are at most $\deg_YQ$ such pairs.

 For the second, we will need a nonzero derivative in order to
 perform Hensel lifting. Specifically, once $\alpha,\beta$ are fixed,
 we use the known equation $Q(X,\alpha+\beta X,P(X))=0$, together with
 one value of $P$, to recover the coefficients of $P$ one at a time.
 Each step requires dividing by the $Z$-derivative at the starting root. Thus we
 need to start at a point where this derivative is nonzero, which we
 can choose when $Q_Z(X,\alpha+\beta X,P(X))\not\equiv0$.
 To count this second case, observe that $P$ is a common root of
 $Q(X,\alpha+\beta X,Z)$ and its $Z$-derivative. Hence their resultant
 vanishes along the challenge line:
 \[
  \left.\Res_Z(Q,Q_Z)\right|_{Y=\alpha+\beta X}\equiv0.
 \]
 This resultant is nonzero because $Q$ is irreducible and $Q_Z$ is
 nonzero of smaller $Z$-degree. Each such pair therefore gives a
 linear factor $Y-\alpha-\beta X$ of the resultant, and we bound their
 number by its $Y$-degree, just as in the first case. This also counts
 challenges at which the restriction has any repeated root, whether
 or not that root is a polynomial $P$.

 Straightforward degree counts then bound the total number of excluded
 pairs by $D^2$, after summing over all factors.
 Here we use that the interpolated explainer satisfies
 $\deg_{Y,Z}Q\le D=O(1/(\rho\eta^2))$, independent of $n$.

 \item \emph{Persistent triples.}
 In this step, we count exceptional triples whose agreement can be explained by correlated agreement with a pair of rational functions. That is, $(\alpha, \beta, P)$ such that we can write
 \[
  P=F+(\alpha+\beta X)G,
 \]
 where $F,G$ have a common denominator of degree at most one and
 numerators of degree at most $d+1$ and $d$, respectively, and have
 correlated agreement with $f,g$ on a large set $I\subseteq\calD$.
 Such an explanation need not be an MCA witness. 

 Applying standard list-decoding up to the Johnson bound to such
 rational pairs gives at most $4/\eta$ pairs of rational functions that can have correlated agreement with the received words. For each fixed
pair of rational functions $F,G$, a straightforward counting argument bounds the number of exceptional coefficient pairs
 $(\alpha,\beta)$ associated with it by $\max\{\binom n2,q\}$. This gives an MCA error of
 \[
  \frac{4}{\eta q^2}\max\{\tbinom n2,q\}
  =O_\eta(1/q+n^2/q^2),
 \]
coming from persistent triples.
 \item \emph{Nonpersistent triples.}
 We count these triples by covering them with algebraic sets of
 bounded degree and dimension at most two, then counting agreement
 incidences on these sets. We use standard algebraic geometry and
 adapt the agreement-incidence counting argument of~\cite{DKT} to bound the
 number of nonpersistent triples by $O_{\rho,\eta}(n^2)$. In this step, the nonzero derivative guaranteed (from being outside the excluded
 case) lets us construct the cover by Hensel lifting, and the absence
 of a rational explanation $(F,G)$ as in the persistent case leads to the bound on the dimension.
\end{itemize}
Altogether, this gives MCA error at most $\frac4{\eta q}+O_{\rho,\eta}(n^2/q^2).$

\paragraph{Where the gain in the proof is.}
Here we give some more detail on where in the proof affine combiners
have an advantage over scalar combiners. At a high level, prior proximity gaps results \cite{BCIKS, BCHKS, DKT} follow the same partition when
counting exceptional candidates. The error due to persistent and
nonpersistent candidates is $O(n/q)$ for each. In our case, both give an $O(n^2/q^2)$ term, with an
additional $O(1/q)$ term for persistent candidates.

\emph{Persistent triples.}
Here the gain of affine combiners is relatively easy to see, and one can consider the following toy scenario for simplicity. Suppose $f,g$ have correlated agreement with polynomials $F,G$ of
degrees at most $d,d-1$, respectively, and let $I$ be their full
common agreement domain. For every challenge $(\alpha,\beta)$,
the polynomial
\[
 P_{\alpha,\beta}(X)=F(X)+(\alpha+\beta X)G(X)
\]
agrees with the combined word on $I$. An agreement at a coordinate
$x\in\calD\setminus I$ requires cancellation of the discrepancies, so
\[
 (F(x)-f(x))+(\alpha+\beta x)(G(x)-g(x))=0.
\]
We call this an accidental agreement. With one coordinate of agreement
loss, an exceptional candidate needs at least two such agreements and at any accidental agreement we have $g(x)\ne G(x)$, so writing
$r(x)=(F(x)-f(x))/(g(x)-G(x))$, two distinct such coordinates give
\[
 \alpha+\beta x_1=r(x_1),\qquad \alpha+\beta x_2=r(x_2).
\]
These equations determine exactly one $(\alpha,\beta)$. Taking a union
bound over pairs of coordinates outside $I$ gives
\[
 \Pr_{\alpha,\beta}[\text{at least two accidental agreements}]
 \le \frac{\binom{n-|I|}{2}}{q^2}.
\]

For scalar mixing, the same two coordinates give the equations
$\alpha=r(x_1)$ and $\alpha=r(x_2)$. We can arrange the coordinates
outside $I$ into disjoint pairs and choose the values of $f,g$ so that
the two equations in each pair have the same solution, with distinct
solutions across pairs. This gives $\floor{(n-|I|)/2}$ scalar challenges
with two accidental agreements, out of $q$ possible challenges. The
construction in \cref{sec:separation} makes such challenges exceptional
and gives the scalar $\Omega(n/q)$ lower bound.

\emph{Nonpersistent triples.}
Here the second challenge coefficient $\beta$ gives us $q^2$ possible
challenges and increases the dimension of the algebraic sets used in
the count. We cover the triples $(\alpha,\beta,P)$ by algebraic sets of
dimension at most two and total degree $O(n^2)$. The scalar proof of
Dao, Kominers, and Thaler~\cite{DKT} covers pairs
$(\alpha,P)$ by algebraic sets of dimension at most one and total
degree $O(n)$. Ultimately, this gives
$O(n^2/q^2)$ error for affine combiners and $O(n/q)$ error for scalar
combiners.


\section{Preliminaries}\label{sec:event}
Throughout this paper, we work over a finite field $\F_q$, and fix a size $n$ evaluation domain $\calD \subseteq \F_q$. Let $d \in \mathbb{N}$ be a degree bound and set $\rho = d/n$ as the fractional rate along with the following parameters
\begin{equation}\label{eq:parameters}
\begin{gathered}
 \sqrt{\frac{4}{d}}\le\eta<1-\sqrt\rho, \qquad
 A\coloneq\ceil{(\sqrt\rho+\eta)n}.
\end{gathered}
\end{equation}
Above, $\eta$ is a slack parameter and $A$ is an absolute agreement parameter. We fix two received words $f,g:\calD\to\F_q$.
When working with proximity gaps, it will be helpful to define the notion of exceptional triples $(\alpha, \beta, P)$ which are essentially triples where $\deg(P) \leq d$ and $P$ and $w_{\alpha, \beta}$ agree on at least $A$ points over $\mc{D}$, but this agreement is not explained by correlated agreement between $f,g$ with degree $d,d-1$ polynomials.

\begin{definition}
A triple $(\alpha,\beta,P)\in\F_q^2\times\F_q[X]$ is
\emph{exceptional} if $\deg P\le d$, $|S_{\alpha,\beta}(P)|\ge A$,
and $P$ has no admissible witnesses with agreement loss at most $1$,
as defined in Section~\ref{sec:our-results}.
\end{definition}
In this language, bounding the MCA error $\Err_{\rm aff}$ from Theorem~\ref{thm:main}
is equivalent to bounding the number of pairs
$(\alpha,\beta)$ that can occur in an exceptional triple. The proof in the remainder of the paper will take this viewpoint.

We will require the following form of the Johnson bound.
\begin{lemma}[{\cite[Ch.~17, Theorem~3]{MS77}}]\label{lem:johnson}
Let $0<a\le1$ and $0\le b<a^2$. Any family of subsets of an
$n$-element set, each of size at least $an$ and with pairwise
intersections of size at most $bn$, has at most
$\frac{a-b}{a^2-b}$ members.
\end{lemma}

\subsection{Algebraic Geometry}\label{sec:algebraic-geometry}
We recall the algebraic notions used in the root-counting argument.
Rational maps will give covers of polynomial roots by algebraic sets of
bounded dimension and degree, and hyperplane sections will let us count
their agreement incidences. Let $\mathbb{K}$ be an algebraically closed
field containing $\F_q$.

\begin{definition}
An \emph{algebraic set} in $\mathbb{K}^k$ is a common zero set
\[
 V=\{x\in \mathbb{K}^k:F_1(x)=\cdots=F_s(x)=0\},
 \qquad F_1,\ldots,F_s\in \mathbb{K}[X_1,\ldots,X_k].
\]
A \emph{hypersurface} is the zero set of one nonconstant polynomial;
a \emph{hyperplane} is the zero set of a nonconstant affine linear
polynomial. A nonempty algebraic set is \emph{irreducible} if it
cannot be written as the union of two proper algebraic subsets.
Every algebraic set is a finite union of its maximal irreducible
algebraic subsets, called its \emph{irreducible components}.
\end{definition}

\begin{definition}
The \emph{Zariski closure} of $U\subseteq \mathbb{K}^k$, denoted $\overline U$,
is the smallest algebraic set containing $U$.
\end{definition}

\begin{definition}
For a nonempty algebraic set $V$, define
\[
 \dim V\coloneq\max\{r:C_0\subsetneq C_1\subsetneq\cdots\subsetneq C_r\subseteq V\},
\]
where the $C_i$ are nonempty irreducible algebraic sets.
The dimension of $V$ is the largest dimension of its irreducible
components.
We set $\dim\varnothing\coloneq-1$.
\end{definition}

\begin{definition}
For an irreducible algebraic set $C\subseteq \mathbb{K}^k$ of dimension $r$,
let $L\subseteq \mathbb{K}^k$ be a generic affine linear subspace of dimension
$k-r$, and define
\[
 \deg C\coloneq|C\cap L|\qquad\text{(counted with multiplicity)}.
\]
Here \emph{generic} means that $L$ is chosen outside a fixed proper
algebraic subset of the parameter space of such subspaces.
For an algebraic set with irreducible components $C_1,\ldots,C_s$, define
\[
 \deg V\coloneq\sum_{i=1}^s\deg C_i,
\]
including components of smaller dimension. A zero-dimensional
algebraic set is finite and satisfies $\deg V=|V|$.
We set $\deg\varnothing\coloneq0$.
\end{definition}

\begin{definition}
Let $V\subseteq \mathbb{K}^k$ be algebraic and let
$N_1,\ldots,N_t,S\in \mathbb{K}[X_1,\ldots,X_k]$.
On $U\coloneq\{x\in V:S(x)\ne0\}$, these polynomials define a
\emph{rational map}
\[
 \phi:U\longrightarrow \mathbb{K}^t,\qquad
 x\longmapsto\left(\frac{N_1(x)}{S(x)},\ldots,\frac{N_t(x)}{S(x)}\right).
\]
Its \emph{image closure} is $\overline{\phi(U)}$.
\end{definition}

The next two lemmas are standard facts from algebraic geometry.

\begin{lemma}\label{lem:hypersurface}
A hypersurface in $\mathbb{K}^k$ defined by a polynomial of degree $b\ge1$
has dimension $k-1$ and total degree at most $b$.
\end{lemma}

\begin{lemma}\label{lem:hyperplane-section}
If $C\subseteq \mathbb{K}^k$ is an irreducible algebraic set and a hyperplane
$H\subseteq \mathbb{K}^k$ does not contain $C$, then
\begin{equation}\label{eq:hyperplane-section}
 \dim(C\cap H)\le\dim C-1,\qquad
 \deg(C\cap H)\le\deg C.
\end{equation}
\end{lemma}
\begin{proof}
Since $C\cap H$ is a proper closed subset of the irreducible set $C$,
its dimension is strictly smaller than $\dim C$.
The degree bound is the hyperplane case of B\'ezout's
inequality~\cite[Theorem~1]{Hei83}:
$\deg(C\cap H)\le\deg C\cdot\deg H=\deg C$.
\end{proof}

\begin{lemma}[{\cite[Lemma~A.4]{DKT}}]\label{lem:rational-image}
Let $V$ be an algebraic set of dimension at most $r\ge0$, and let
$\phi\coloneq(N_1/S,\ldots,N_t/S)$, where all numerators and the common
denominator have degree at most $b\ge1$. If $S$ is not identically
zero on any irreducible component of $V$, the image closure
$W\coloneq\overline{\phi(\{x\in V:S(x)\ne0\})}$ satisfies
\begin{equation}\label{eq:rational-image}
 \dim W\le r,\qquad \deg W\le(\deg V)b^r.
\end{equation}
\end{lemma}
Components contained in $S=0$ can first be discarded without
increasing the dimension or degree of $V$.

\paragraph{Hensel lifting.}\label{sec:hensel-lifting}
Suppose we have a polynomial equation $R(X,\alpha+\beta X,Z)=0$ and at a particular setting, $X = t_0$, it has a root at $Z = z$, i.e.\ $R(t_0,\alpha+\beta t_0,z)=0$. Then, if the derivative of $R$ with respect to $Z$ is nonzero at
$(t_0,\alpha+\beta t_0,z)$, Hensel lifting constructs a formal power
series $z+E(X-t_0)$ with $E(0)=0$ that solves the equation and captures
its polynomial roots with initial value $z$. More precisely, if a
polynomial $P$ satisfies $R(X,\alpha+\beta X,P(X))=0$ and $P(t_0)=z$,
then
\[
 z+E(X-t_0)=P(X).
\]
The following lemma expresses the coefficients of $E$ as rational functions of
$\alpha,\beta,z$ and bounds their degrees.
\begin{lemma}\label{lem:hensel-lifting}
Let $\mathbb{K}$ be a field, let $R\in\mathbb{K}[X,Y,Z]$, and fix
$t_0,\alpha,\beta,z\in\mathbb{K}$. Suppose $z$ is a root of
$R(t_0,\alpha+\beta t_0,Z)=0$ and $R_Z(t_0,\alpha+\beta t_0,z)\ne0$,
where $R_Z=\partial R/\partial Z$. Then there is a unique formal power
series $z+E(X-t_0)$ with $E(0)=0$ satisfying
\begin{equation}\label{eq:hensel-identity}
 R(X,\alpha+\beta X,z+E(X-t_0))=0
\end{equation}
In particular, any polynomial root $P(X)$ with $P(t_0)=z$ equals
$z+E(X-t_0)$.

Moreover, writing $E(X-t_0)=\sum_{j\ge1}c_j(\alpha,\beta,z)(X-t_0)^j$, we can express each coefficient in the power series as 
\begin{equation}\label{eq:hensel-coefficients}
 c_j(\alpha,\beta,z)=\frac{N_j(\alpha,\beta,z)}{S(\alpha,\beta,z)^{2j-1}},
\end{equation}
where $S(Y,V,Z)=R_Z(t_0,Y+Vt_0,Z)$, $h=\deg_{Y,Z}R$, and
$N_j\in\mathbb{K}[Y,V,Z]$ has total degree at most $1+(2j-1)(h-1)$
for all $j\ge 1$.
\end{lemma}
The proof essentially follows~\cite[Claim~A.2]{BCIKS},
with the notation adapted to the affine term in the second variable.
We include a brief proof for completeness.
\begin{proof}
Set $W=X-t_0$. The polynomial $R(t_0+W,\alpha+\beta(t_0+W),z+U)$
vanishes at $W=U=0$ and its $U$-derivative there is nonzero by assumption.
Then Hensel lifting over $\mathbb{K}$ (see~\cite[Lemma~2.7]{KTZ}) gives
existence and uniqueness. If $P(X)$ is a polynomial root with $P(t_0)=z$,
then $P(X)-z$ has zero constant term in $X-t_0$ and satisfies
\eqref{eq:hensel-identity} in place of $E(X-t_0)$. Uniqueness gives
$P(X)=z+E(X-t_0)$.

For the coefficient bounds, we take the given identity $R(X,\alpha+\beta X,Z)=0$, shift the variable $Z=z+U$, and write out the expansion in these variables as
\[
 R(t_0+W,\alpha+\beta(t_0+W),z+U)
 =\sum_{a,b\ge0}A_{a,b}(\alpha,\beta,z)W^aU^b,
\]
where $A_{a,b}\in\mathbb{K}[Y,V,Z]$ is the coefficient of $W^aU^b$ in
$R(t_0+W,Y+V(t_0+W),Z+U)$. Since $h=\deg_{Y,Z}R$, we have $\deg A_{a,b}\le h-b$.
The coefficient of $U$ at $W=0$ in this expansion is
$S(\alpha,\beta,z)$, and $\deg S\le h-1$. We write $S$ for
$S(\alpha,\beta,z)$ in the coefficient comparison below.

Write $E(W)=\sum_{j\ge1}c_jW^j$. Substituting $U=E(W)$ into the expansion gives
\[
 0=\sum_{a,b\ge0}A_{a,b}(\alpha,\beta,z)W^a
       \left(\sum_{j\ge1}c_jW^j\right)^b.
\]
To prove~\eqref{eq:hensel-coefficients}, we compare coefficients of $W^j$ on both sides of the equation one by one to sequentially ``reconstruct'' all of them. Then, the degree bound on $N_j$ can be proved inductively.

First, for $j=1$, the coefficient of $W$ on the right handside must cancel, so we get $A_{1,0}(\alpha, \beta, z) + Sc_1 = 0$  and
\[ 
c_1=-\frac{A_{1,0}(\alpha,\beta,z)}{S}.
\]
Then, the $j=1$ case of~\eqref{eq:hensel-coefficients} holds with $N_1=-A_{1,0}$,and indeed $\deg N_1\le h$.

For $j\ge2$, the coefficient of $W^j$ is $Sc_j$ plus terms of the form
\[
 A_{a,b}c_{j_1}\cdots c_{j_b},\qquad
 a+j_1+\cdots+j_b=j,\quad 1\le j_1,\ldots,j_b<j,
\]
$c_j$ is $-1/S$ times the sum of these terms. By induction, we can write each $c_{j_\ell}$ in the form $N_{j_\ell}/S^{2j_\ell-1}$.
$c_{j_\ell}=N_{j_\ell}/S^{2j_\ell-1}$ where $N_{j_\ell}$ satisfies the degree bound in~\eqref{eq:hensel-coefficients}. Substituting and using
$j_1+\cdots+j_b=j-a$, each summand of $c_j$ becomes
\[
\begin{aligned}
 -\frac{1}{S}\cdot A_{a,b}\prod_{\ell=1}^b\frac{N_{j_\ell}}{S^{2j_\ell-1}}
 &= -\frac{1}{S}\cdot\frac{A_{a,b}N_{j_1}\cdots N_{j_b}}{S^{2(j-a)-b}}\\
 &= -\frac{1}{S}\cdot\frac{A_{a,b}S^{2a+b-2}N_{j_1}\cdots N_{j_b}}{S^{2j-2}}.
\end{aligned}
\]
If $a=0$, then $b\ge2$ since the indices $j_\ell$ sum to $j$ and
each is less than $j$. Thus $2a+b-2\ge0$ and each exponent of $S$ is positive. This shows that we can write $c_j$ with denominator $S^{2j-1}$ as desired. Then $N_j$ is the sum of the numerators above and it remains to check its degree bound. Using $\deg A_{a,b}\le h-b$,
$\deg S\le h-1$, and the inductive bounds, each numerator satisfies
\[
 \deg\!\left(A_{a,b}S^{2a+b-2}N_{j_1}\cdots N_{j_b}\right)
 \le h+(2j-2)(h-1)=1+(2j-1)(h-1)
\]
as desired
\end{proof}

\section{Proof of the proximity gap for affine combiners}\label{sec:main-proof}
We now prove \Cref{thm:main} by bounding the number of challenges $(\alpha, \beta)$ which are part of an exceptional triple.

\begin{lemma}\label{lem:counting}
For every $f,g:\calD\to\F_q$, the number of pairs
$(\alpha,\beta)\in\F_q^2$ for which $(\alpha,\beta,P)$ is exceptional
for some $P$ is at most
\[
 \frac{4q}{\eta}+\frac{2^{14}n^2}{\rho^3\eta^{10}}.
\]
\end{lemma}
Dividing the bound above by $q^2$ proves \Cref{thm:main}. The remainder of this section is dedicated to the proof of
Lemma~\ref{lem:counting}. We follow the interpolation and root-counting
approach of Guruswami and Sudan~\cite{GS} for list decoding, extended
to proximity gaps by Ben-Sasson, Carmon, Ishai, Kopparty, and
Saraf~\cite{BCIKS}.

Specifically, from $f$ and $g$, we interpolate a nonzero polynomial
$Q\in\F_q[X,Y,Z]$ such that every exceptional triple
$(\alpha,\beta,P)$ satisfies
\[
 Q(X,\alpha+\beta X,P(X))=0
 \qquad\text{in }\F_q[X].
\]
Once $Q$ is fixed, to prove Lemma~\ref{lem:counting}, we must bound the number of coefficient pairs $(\alpha,\beta)$ for which this there exists $P$ forming an exceptional triple with it.

\subsection{Interpolation along affine lines}\label{sec:interp}
We construct one polynomial relation satisfied by every candidate with
at least $A$ agreements. For $H(X,Y,Z)=\sum_{i,j,k}h_{ijk}X^iY^jZ^k\ne0$,
define its $(1,1,d)$-weighted degree by
\[
 \deg_{1,1,d}(H)\coloneq\max_{h_{ijk}\ne0}(i+j+dk).
\]

For the interpolation, define
\[
 m\coloneq\ceil{2/\eta},\qquad
 D\coloneq\ceil{\frac{(m+1/2)^2}{3\rho}}-1,
\]
and
\begin{equation}\label{eq:param}
 M\coloneq\min\left\{D,\floor{\frac{mA-D-1}{d}}\right\}.
\end{equation}

\begin{lemma}\label{lem:roots}
For every pair of received words $f,g:\calD\to\F_q$, there exists a
nonzero polynomial $Q\in\F_q[X,Y,Z]$ such that:
\begin{itemize}
 \item $\deg_{Y,Z}Q\le D$ and $1\le\deg_ZQ\le M$.
 \item $\deg_{1,1,d}Q<mA$.
 \item For every $\alpha,\beta\in\F_q$ and every
 $P\in\F_q[X]$ with $\deg P\le d$ and
 $|S_{\alpha,\beta}(P)|\ge A$,
 \begin{equation}\label{eq:lineidentity}
  Q(X,\alpha+\beta X,P(X))=0.
 \end{equation}
\end{itemize}
\end{lemma}
\begin{proof}
We adapt the interpolation argument of~\cite[Lemma~3.1]{BCHKS}.
Consider polynomials of the form
\[
 Q(X,Y,Z)=\sum_{\substack{i,t,j\ge0\\t+j\le D,\ i+dj<mA-D}}
 Q_{i,t,j}X^iY^tZ^j,
\]
where the coefficients $Q_{i,t,j}\in\F_q$ are treated as unknowns.
Each monomial has $j\le M$, $t+j\le D$, and $i+t+dj<mA$,
so the required degree upper bounds hold for all polynomials we are considering.
Let us view $Q$ as a polynomial in $X,Z$ over $\F_q[Y]$.
Then, for each $z\in\calD$, we require that $Q$ has a root of multiplicity at least $m$ at
$(z,f(z)+Yg(z))$. Specifically, for every $r,s\ge0$ with $r+s<m$, impose
\begin{equation}\label{eq:graphmultiplicity}
 \sum_{\substack{i\ge r,\ t\ge0,\ j\ge s\\
                  t+j\le D,\ i+dj<mA-D}}
 Q_{i,t,j}\binom{i}{r}\binom{j}{s}
 z^{i-r}Y^t\bigl(f(z)+Yg(z)\bigr)^{j-s}=0
\end{equation}
where the equation is over $\F_q[Y]$.

Each summand has $Y$-degree at most $t+j-s\le D-s$.
Thus, in~\eqref{eq:graphmultiplicity}, we impose at most $D-s+1$ homogeneous
linear equations over $\F_q$, one for each coefficient of $Y$.
For a fixed $s<m$, there are $m-s$ choices of $r$ and $n$ choices of
$z$; if $s>M$, the sum is empty. The total number of scalar equations
is therefore at most
\begin{equation} \label{eq:unknowns}
 n\sum_{s=0}^{\min(m-1,M)}(m-s)(D-s+1).
\end{equation}

Now we show that any $Q$ satisfying these constraints satisfies the vanishing
condition in the third item above. Suppose $z \in \mc{D}$ is a point where
$P(z) = w_{\alpha,\beta}(z)$ for some $\alpha,\beta \in \F_q$.
The multiplicity constraints imply that $Q$ is a sum of polynomial multiples of
\[
 (X-z)^r\bigl(Z-f(z)-Yg(z)\bigr)^s,\qquad r+s=m.
\]
After substituting $Y=\alpha+\beta X$ and $Z=P(X)$, the expression
$Z-f(z)-Yg(z)$ becomes $P(X)-f(z)-(\alpha+\beta X)g(z)$ which is divisible by $X-z$. Thus each summand in \eqref{eq:graphmultiplicity} is divisible by $(X-z)^m$, and $z$ is a root of
multiplicity at least $m$ of $Q(X,\alpha+\beta X,P(X))$.

If $P$ agrees with $w_{\alpha,\beta}$ on at least $A$ points, we obtain
at least $mA$ roots counted with multiplicity. On the other hand,
$\deg P\le d$ and the weighted-degree bound on $Q$ give
\[
 \deg Q(X,\alpha+\beta X,P(X))
 \le\deg_{1,1,d}Q<mA.
\]
Therefore $Q(X,\alpha+\beta X,P(X))$ vanishes identically as required by the third item above.

It remains to show that the constraints have a nonzero solution.
For each $0\le j\le M$, there are $mA-D-dj$ choices of $i$ and
$D-j+1$ choices of $t$. The number of unknowns is therefore
\[
 \sum_{j=0}^M(mA-D-dj)(D-j+1).
\]

Set $U\coloneq\ceil{(m+1/2)\sqrt{nd}}$.
A direct calculation, as in the proof of~\cite[Lemma~3.1,
equations~(11)--(12)]{BCHKS} with their $k=d$, gives
\[
 \sum_{j=0}^{\floor{(U-1)/d}}(U-dj)(D-j+1)
 >n\sum_{s=0}^{m-1}(m-s)(D-s+1),
\]
where $m\le\floor{(U-1)/d}\le D$.
By our choice of $m$, we have $m\ge3$ and $2\le m\eta<3$, so by~\eqref{eq:parameters},
\[
 D+1\le\frac{(m+1/2)^2}{3\rho}+1
 \le\frac{2m^2}{3\rho}\le\frac{m\eta n}{2}.
\]
Consequently,
\[
 mA-U\ge(m\eta-\tfrac12\sqrt\rho)n-1
 \ge\frac{m\eta n}{2}-1\ge D.
\]
Hence $U\le mA-D$ and $m\le\floor{(U-1)/d}\le M$, and therefore
\begin{equation}\label{eq:dimension}
 \sum_{j=0}^M(mA-D-dj)(D-j+1)
 >n\sum_{s=0}^{\min(m-1,M)}(m-s)(D-s+1).
\end{equation}
There are therefore more unknowns than equations, whose number is bounded
by~\eqref{eq:unknowns}, so a nonzero solution $Q$ exists.

Fix such a $Q$. The last step is to show that $\deg_ZQ\ge1$. To see this, note that if $\deg_ZQ=0$, then
$(X-z)^m$ divides $Q$ for every $z\in\calD$ by the multiplicity
conditions, and this contradicts $\deg_XQ<mA\le mn$. Thus $\deg_ZQ\ge1$ and the polynomial $Q$ satisfies all of the conditions of the lemma.
\end{proof}

We fix a polynomial $Q$ provided by \cref{lem:roots} for the rest of this section. As mentioned, our goal henceforth reduces to a root counting problem on $Q$ and we now describe in more detail how weo do so using a partitioning of the. Specifically, we will want to bound the number of coefficient pairs $(\alpha, \beta)$ such that there exists a low degree $P$ satisfying $Q(X, \alpha + \beta X, P(X)) =0$ and $(\alpha, \beta, P)$ is exceptional.

We factor $Q$ as
\[
 Q=C(X,Y)\prod_{i=1}^r Q_i(X,Y,Z)^{a_i},
\]
where the $Q_i$ are distinct irreducible factors of positive
$Z$-degree. For each $i$, let $e_i$ be the largest power of the field
characteristic dividing every positive $Z$-exponent of $Q_i$. Note that we can find $R_i$ such that 
\[
 Q_i(X^{e_i},Y^{e_i},Z)=R_i(X,Y,Z)^{e_i}.
\]
Henceforth, we fix these $Q_i$, $R_i$, and $e_i$. Define
$h_i\coloneq\deg_{Y,Z}R_i$. Note that
\begin{equation}\label{eq:factor-bounds}
 h_i\ge1,\qquad e_i\le M,\qquad \deg_YC+\sum_i h_i\le D.
\end{equation}
The first bound is trivial, the second bound is because  $e_i$ divides every positive $Z$-exponent of $Q_i$ and $Q$ has $Z$-degree at most $M$. The third bound is because the $Y,Z$ degree of $Q$ is at most $D$.

Denote $\partial R_i/\partial Z$ by $R_{i,Z}$.
If $Q_i(X,\alpha+\beta X,P(X))=0$, then, after substituting $X^{e_i}$ for $X$ we can obtain
\begin{equation}\label{eq:normalizedline}
 R_i\bigl(X,\alpha^{1/e_i}+\beta^{1/e_i} X,P(X^{e_i})\bigr)=0.
\end{equation}

Now, define the excluded coefficient pairs by
\begin{equation}\label{eq:excluded-coefficients}
 \mathcal E\coloneq\left\{(\alpha,\beta)\in\F_q^2:
 \begin{array}{l}
 C(X,\alpha+\beta X)\equiv0,\text{ or}\\[2pt]
 R_i(X,\alpha^{1/e_i}+\beta^{1/e_i}X,Z)\text{ has a repeated}\\
 \text{root in }Z\text{ for some }i
 \end{array}\right\},
\end{equation}
where, above, roots are taken over $\overline{\F_q(X)}$.

Set the intermediate agreement threshold
\[
 B\coloneq\ceil{(\sqrt\rho+\eta/2)n}-1.
\]
We call a triple $(\alpha,\beta,P)\in\F_q^2\times\F_q[X]$, with
$\deg P\le d$, \emph{persistent} if its agreement on a set of
at least $B$ coordinates has a ``rational explanation'' in the following sense. There exist a set
$I\subseteq S_{\alpha,\beta}(P)$ with $|I|\ge B$, rational functions
$F,G\in\F_q(X)$ such that $P = F + (\alpha + \beta X) G$ and
\[
\begin{gathered}
 F(z)=f(z)\text{ and }G(z)=g(z)\quad\text{for all }z\in I.
\end{gathered}
\]
where $G = H/L$, $\deg H \le d$, $L \ne 0$, and $\deg L \le 1$. We say that $F, G$ \emph{explains} the persistence of $(\alpha,\beta,P)$.

We call the triple \emph{nonpersistent} otherwise. Denote the persistent and
nonpersistent exceptional coefficient pairs outside $\mathcal E$ by
\begin{equation}\label{eq:persistent-coefficients}
 \mathcal P\coloneq\left\{(\alpha,\beta)\in\F_q^2\setminus\mathcal E:
 \begin{array}{l}
 \exists P\text{ such that }(\alpha,\beta,P)\text{ is exceptional}\\
 \text{and persistent}
 \end{array}\right\},
\end{equation}
\begin{equation}\label{eq:nonpersistent-coefficients}
 \mathcal N\coloneq\left\{(\alpha,\beta)\in\F_q^2\setminus\mathcal E:
 \begin{array}{l}
 \exists P\text{ such that }(\alpha,\beta,P)\text{ is exceptional}\\
 \text{and nonpersistent}
 \end{array}\right\}.
\end{equation}
Every coefficient pair occurring in an exceptional triple lies in
$\mathcal E\cup\mathcal P\cup\mathcal N$ hence it suffices for us to bound the sizes of each of these three sets. 

\subsection{Bounding excluded coefficient pairs}\label{sec:excluded}
We first bound $|\mathcal E|$ and also show a fact about roots outside of $\mc{E}$ which will be useful later.

\begin{lemma}\label{lem:excluded}
We have
\[
 |\mathcal E|\le D^2.
\]
Furthermore, for every $(\alpha,\beta)\notin\mathcal E$ and every
$P\in\F_q[X]$ of degree at most $d$ satisfying
\[
 Q(X,\alpha+\beta X,P(X))=0,
\]
there is some $i\in[r]$ such that
\[
\begin{aligned}
 R_i\bigl(X,\alpha^{1/e_i}+\beta^{1/e_i} X,P(X^{e_i})\bigr)&=0,\\
 R_{i,Z}\bigl(X,\alpha^{1/e_i}+\beta^{1/e_i} X,P(X^{e_i})\bigr)&\not\equiv0.
\end{aligned}
\]
\end{lemma}
\begin{proof}
For any nonzero $G(X,Y)$,
\[
 G(X,\alpha+\beta X)=0
 \quad\Longleftrightarrow\quad
 Y-\alpha-\beta X\text{ divides }G(X,Y).
\]
Distinct pairs correspond to distinct factors of $Y$-degree one,
so the restriction vanishes identically for at most $\deg_YG$ pairs.
Applying this to $C$, at most $\deg_YC$ pairs satisfy the first
condition defining $\mathcal E$.

For the second condition, fix $i$. The specialization
$R_i(X,\alpha^{1/e_i}+\beta^{1/e_i}X,Z)$ is nonzero as otherwise
$Y-\alpha^{1/e_i}-\beta^{1/e_i}X$ would divide the irreducible
polynomial $R_i$, which has positive $Z$-degree.
If the specialization has a repeated root, then its $Z$-derivative has the same root. By the polynomial B\'ezout identity,
\[
 \left.\Res_Z(R_i,R_{i,Z})\right|_{Y=\alpha^{1/e_i}+\beta^{1/e_i}X}
 \equiv0.
\]
This resultant is nonzero because $R_i$ is irreducible and
$R_{i,Z}\ne0$. Since $R_i$ and $R_{i,Z}$ have total $(Y,Z)$-degrees
at most $h_i$ and $h_i-1$ respectively, the Sylvester determinant has 
$Y$-degree at most $h_i(h_i-1)$.
Hence, there are at most $h_i(h_i-1)$ pairs satisfying the second condition for this $i$ 
and, summing over all $i$, we obtain
\[
\begin{aligned}
 |\mathcal E|&\le\deg_YC+\sum_i h_i(h_i-1)\\
 &\le\left(\deg_YC+\sum_i h_i\right)^2\le D^2,
\end{aligned}
\]
where the last inequality uses~\eqref{eq:factor-bounds}. This completes the bound on $|\mc{E}|$.

We move on to the second part of the lemma. Fix $(\alpha,\beta)\notin\mathcal E$ and a polynomial $P$
satisfying~\eqref{eq:lineidentity}. Substituting into the factorization of
$Q$, we obtain
\[
 0=Q(X,\alpha+\beta X,P(X))
  =C(X,\alpha+\beta X)\prod_i Q_i(X,\alpha+\beta X,P(X))^{a_i}.
\]
Since $(\alpha,\beta)\notin\mathcal E$,
$C(X,\alpha+\beta X)$ is nonzero and hence for at least one $i$ we have
\[
 R_i\bigl(X,\alpha^{1/e_i}+\beta^{1/e_i}X,P(X^{e_i})\bigr)\equiv0,
\]
and moreover 
$R_i(X,\alpha^{1/e_i}+\beta^{1/e_i}X,Z)$ has no repeated roots in $Z$.
This means its root $Z=P(X^{e_i})$ is simple and cannot also be a root
of its $Z$-derivative, so
\[
 R_{i,Z}\bigl(X,\alpha^{1/e_i}+\beta^{1/e_i}X,P(X^{e_i})\bigr)
 \not\equiv0.
\]
\end{proof}

\subsection{Bounding persistent coefficient pairs}\label{sec:persistent}
Towards bounding $|\mathcal P|$, we first need the following fact.
It bounds the number of exceptional coefficient pairs compatible with
one fixed rational pair $F,G$.

\begin{lemma}\label{lem:rational-pair}
For every pair of rational functions $F, G\in\F_q(X)$, the number of coefficient pairs
$(\alpha,\beta)\in\F_q^2$ for which $P\coloneq F+(\alpha+\beta X)G$
is a polynomial and $(\alpha,\beta,P)$ is exceptional is at most
\[
 \begin{cases}
 \binom n2,&\text{ if } F,G\in\F_q[X],\ \deg F\le d,\ \deg G\le d-1,\\
 q,&\text{otherwise}.
 \end{cases}
\]
\end{lemma}
\begin{proof}
If $F,G$ are not both polynomials with $\deg F\le d$ and
$\deg G\le d-1$, the pairs $(\alpha,\beta)$ for which
$F+(\alpha+\beta X)G$ is a polynomial of degree at most $d$ form
an affine subspace of $\F_q^2$, if nonempty. This subspace cannot contain $\F_q^2$ as 
otherwise the pairs $(0,0),(1,0),(0,1)$ would force $F,G,XG$ to all be
polynomials of degree at most $d$, which is the opposite of what we assumed. There are thus at most $q$ pairs in this case, taking care of the second case above.

Now suppose $F,G\in\F_q[X]$ with $\deg F\le d$ and
$\deg G\le d-1$. Fix
\[
 I'\coloneq\{z\in\calD:F(z)=f(z)\text{ and }G(z)=g(z)\},
\]
to be the set of points where $F, G$ have correlated agreement with $f, g$ respectively.

Every exceptional coefficient pair has at least two agreement
coordinates outside $I'$; otherwise $F,G$ would be admissible
witnesses for $P$ with agreement loss at most $1$.

Suppose two coefficient pairs $(\alpha,\beta)$ and
$(\alpha',\beta')$, with candidates $P=F+(\alpha+\beta X)G$ and
$P'=F+(\alpha'+\beta'X)G$, share two distinct agreement coordinates
$z,z'\notin I'$. At either coordinate $x\in\{z,z'\}$, we have
\begin{equation}\label{eq:extra}
 (f(x)-F(x))+(\alpha+\beta x)(g(x)-G(x))=0.
\end{equation}
Here $g(x)\ne G(x)$: otherwise this equation would also give
$f(x)=F(x)$, contradicting $x\notin I'$.
Subtracting the corresponding equation for $(\alpha',\beta')$
and dividing by $g(x)-G(x)$ gives
\[
 (\alpha-\alpha')+(\beta-\beta')z=0,\qquad
 (\alpha-\alpha')+(\beta-\beta')z'=0.
\]
Since $z\ne z'$, these equations imply
$(\alpha,\beta)=(\alpha',\beta')$.
Thus each exceptional coefficient pair has a pair of agreement
coordinates outside $I'$ that no other coefficient pair shares.
There are at most $\binom n2$ choices for this unordered pair of
coordinates, proving the first case of the lemma.
\end{proof}

Define
\[
 \ell\coloneq\frac{n(B-d-1)}{B^2-n(d+1)},
\]
so that $\ell$ is the upper bound from \Cref{lem:johnson}, on the size of any family of subsets of $\calD$,
each of size at least $B$ and with pairwise intersections of size
at most $d+1$.

\begin{lemma}\label{lem:persistent}
We have
\[
 |\mathcal P|\le\ell\max\{q,\tbinom n2\}.
\]
\end{lemma}
\begin{proof}
We first bound the number of possible agreement sets.
For each rational pair $F,G$ explaining the persistence of an
exceptional triple, consider, as in the proof of
Lemma~\ref{lem:rational-pair}, set
\[
 I\coloneq\{z\in\calD:F(z)=f(z)\text{ and }G(z)=g(z)\},
\]
to be set of points where $F, G$ have correlated agreement with $f,g$ respectively.

Let $(F_0,G_0)$ and $(F_1,G_1)$ be two distinct such pairs, with
correlated agreement sets $I_0,I_1$. For each $j=0,1$, choose a
corresponding exceptional triple $(\alpha_j,\beta_j,P_j)$ and a
representation $G_j=H_j/L_j$.
At every $z\in I_0\cap I_1$, we have
\[
 F_0(z)=F_1(z)=f(z),\qquad G_0(z)=G_1(z)=g(z).
\]
If $G_0\ne G_1$, then the polynomial
\[
 H_0L_1-H_1L_0
\]
is nonzero, has degree at most $d+1$, and vanishes on every such $z$.
If $G_0=G_1$, then $F_0\ne F_1$, so the polynomial
\[
 (P_0-P_1)L_0
 -(\alpha_0-\alpha_1+(\beta_0-\beta_1)X)H_0
\]
is nonzero, has degree at most $d+1$, and again vanishes on every such $z$.
In either case, $|I_0\cap I_1|\le d+1$.
Since each correlated agreement set has size at least $B>d+1$, it
determines a unique rational pair among those considered above.
By \Cref{lem:johnson}, with $a=B/n$ and $b=(d+1)/n$,
there are at most $\ell$ distinct correlated agreement sets.

Finally, by \Cref{lem:rational-pair}, the rational pair determined by
each correlated agreement set accounts for at most
$\max\{q,\binom n2\}$ exceptional coefficient pairs. Every coefficient
pair in $\mathcal P$ has an exceptional triple whose persistence is
explained by one of these rational pairs, so
\[
 |\mathcal P|\le\ell\max\{q,\tbinom n2\}.
\]
\end{proof}

\subsection{Bounding nonpersistent coefficient pairs}\label{sec:nonpersistent}\label{sec:root}
We now bound $|\mathcal N|$. It will be helpful to define $v\coloneq\frac{n}{A-B}$. Let $T_0$ be an indeterminate and let $\mathbb{K}\coloneq\overline{\F_q(T_0)}$
be the algebraic closure of $\F_q(T_0)$. Since $T_0$ is transcendental, a nonzero polynomial in $\F_q[X]$
must be nonzero when evaluated at $T_0$.

\begin{definition}
We identify triples $(\alpha,\beta,P)$ with coefficient vectors in
\[
 \mathcal A\coloneq \mathbb{K}^{d+3},\qquad
 (\alpha,\beta,P)\longleftrightarrow
 (\alpha,\beta,p_0,\ldots,p_d),
\]
where $P(X)=\sum_{j=0}^d p_jX^j$.
The triples we count lie in $\F_q^{d+3}\subseteq\mathcal A$;
the algebraic sets containing them are defined over $\mathbb{K}$.
\end{definition}

\begin{lemma}\label{lem:nonpersistent}
We have
\[
 |\mathcal N|\le4d^2M^2D^3v^2.
\]
\end{lemma}

For each $z\in\calD$, define the agreement hyperplane
\[
 \mathcal H_z\coloneq\left\{(\alpha,\beta,p_0,\ldots,p_d)\in\mathcal A:
       \sum_{j=0}^d p_jz^j=f(z)+(\alpha+\beta z)g(z)\right\}.
\]

Our first step is to adapt the induction in~\cite[proof of Lemma~4.1]{DKT}
to the affine case we consider. Given an algebraic set of bounded dimension, we bound the number of nonpersistent triples $(\alpha,\beta,P)$ it contains.
\begin{claim}\label{lem:agreement-incidence}
For every integer $r\ge0$, an algebraic set $V\subseteq\mathcal A$
of dimension at most $r$ contains at most $(\deg V)v^r$ nonpersistent
triples $(\alpha,\beta,P)\in\F_q^{d+3}$ with
$|S_{\alpha,\beta}(P)|\ge A$.
\end{claim}
\begin{proof}
We first show the result for an irreducible component $V_0$.
We proceed by induction on $t\coloneq\dim V_0$.
If $t=0$, then $V_0$ is a point and $\deg V_0=1$, so the bound holds.
Suppose $t>0$, and define
\[
 c\coloneq\bigl|\{z\in\calD:V_0\subseteq\mathcal H_z\}\bigr|,
\]
the number of agreement hyperplanes $\mathcal H_z$ that contain $V_0$.

First consider the case where $c>B$. We show that $V_0$ contains no
nonpersistent $\F_q$-triples. Take any $\F_q$-triple
$(\alpha,\beta,P)\in V_0$, and define
\[
 W\coloneq\bigcap_{z\in\calD:\,V_0\subseteq\mathcal H_z}\mathcal H_z.
\]
Since $V_0\subseteq W$, we have $\dim W\ge t>0$.
The set $W$ is defined by affine linear equations over $\F_q$ and
contains $(\alpha,\beta,P)$, so it must contain a second $\F_q$-triple
$(\alpha',\beta',P')\ne(\alpha,\beta,P)$.
Both triples satisfy the same $c$ agreement equations, so
\[
 P'(z)-P(z)=\bigl((\alpha'-\alpha)+(\beta'-\beta)z\bigr)g(z)
 \qquad\text{whenever }V_0\subseteq\mathcal H_z.
\]
Their coefficient pairs, $(\alpha,\beta)$ and $(\alpha',\beta')$, must differ as otherwise $P'-P$ would have degree
at most $d$ and vanish at $c>d$ coordinates, forcing the triples to be the same. We can then define
\[
 G(X)\coloneq\frac{P'(X)-P(X)}{(\alpha'-\alpha)+(\beta'-\beta)X},
 \qquad F(X)\coloneq P(X)-(\alpha+\beta X)G(X).
\]
The polynomial $L\coloneq(\alpha'-\alpha)+(\beta'-\beta)X$ vanishes on at most one of these $c$ coordinates $z$ where $\mathcal H_z\supseteq V_0$.
At each remaining coordinate, $G(z)=g(z)$ and, since the triple
lies in $\mathcal H_z$, also $F(z)=f(z)$.
Thus $F,G$ have correlated agreement with $f,g$ on at least
$c-1\ge B$ coordinates which makes $(\alpha, \beta, P)$ persistent. Hence $V_0$ contains no
nonpersistent $\F_q$-triples in this case.

Now suppose $c\le B$. Let $\mathcal T$ be the set of nonpersistent
$\F_q$-triples in $V_0$ with at least $A$ agreements. We bound
$|\mathcal T|$ by counting the edges of a bipartite graph. Its left
vertices are the triples in $\mathcal T$, and its right vertices
are the $n-c$ coordinates $z\in\calD$ for which
$V_0\not\subseteq\mathcal H_z$. Add an edge between $(\alpha,\beta,P)$ and $z$ if
\[
 P(z)=f(z)+(\alpha+\beta z)g(z),
\]
or equivalently, if $(\alpha,\beta,P)\in V_0\cap\mathcal H_z$.

Every left vertex has at least $A$ agreement coordinates.
Exactly $c$ of these correspond to hyperplanes containing all of
$V_0$; these coordinates are excluded from the right vertices.
Thus every left vertex has at least $A-c$ neighbors, and the graph
has at least $(A-c)|\mathcal T|$ edges.

For a right vertex $z$, its neighbors are precisely the triples in
$\mathcal T\cap\mathcal H_z$. By \Cref{lem:hyperplane-section},
$V_0\cap\mathcal H_z$ has dimension at most $t-1$ and degree at
most $\deg V_0$. Applying induction to its irreducible components
and summing their bounds, we get that it contains at most
$(\deg V_0)v^{t-1}$ nonpersistent $\F_q$-triples with at least
$A$ agreements. Thus each right vertex has at most this many
neighbors. Since there are $n-c$ right vertices, the graph has
at most $(n-c)(\deg V_0)v^{t-1}$ edges.

Comparing these lower and upper bounds on the number of edges gives
\[
 (A-c)|\mathcal T|\le(n-c)(\deg V_0)v^{t-1}.
\]
Since $A-c\ge A-B>0$ and $v=n/(A-B)$, we obtain
\[
 |\mathcal T|\le\frac{n-c}{A-c}(\deg V_0)v^{t-1}
 \le(\deg V_0)v^t.
\]

This establishes the result for irreducible components. For the full lemma, we 
sum over irreducible components.
\end{proof}

To apply the bound above, we conclude by covering the polynomial roots by
algebraic sets of dimension at most two.

\begin{claim}\label{lem:envelope}
There are algebraic sets $V_i\subseteq\mathcal A$, for $i \in [r]$, such that
every $\F_q$-triple satisfying~\eqref{eq:lineidentity} with
$(\alpha,\beta)\notin\mathcal E$ and $\deg P\le d$ lies in
$\bigcup_i V_i$. Furthermore, for each $i$, we have the following dimension and degree bounds
\[
 \dim V_i\le2,\qquad \deg V_i\le4d^2e_i^2h_i^3.
\]
\end{claim}
\begin{proof}
  We fix an $i \in [r]$ and write $R\coloneq R_i$, $e\coloneq e_i$, $h\coloneq h_i$, and
\begin{equation}\label{eq:factor-root-triples}
 \mathcal R_i\coloneq\left\{(\alpha,\beta,P)\in\F_q^2\times\F_q[X]:
 \begin{array}{l}
 (\alpha,\beta)\notin\mathcal E,\quad \deg P\le d,\\
 Q_i(X,\alpha+\beta X,P(X))\equiv0
 \end{array}\right\}.
\end{equation}
Our goal is to cover $\mathcal R_i$ by an algebraic set of dimension at
most two and bounded degree. Our strategy will be to construct a rational
map of bounded degree whose image contains $\mathcal R_i$ and then use
Lemma~\ref{lem:rational-image} to bound the dimension and degree of the
closure of its image.

To this end, observe that for every triple
$(\alpha,\beta,P)\in\mathcal R_i$, we can actually reconstruct $P$
from $\alpha,\beta$ and a single initial value, $P(T_0^e)$, by Hensel lifting.

Indeed, letting $Y=\alpha^{1/e}$, $V=\beta^{1/e}$, and
$Z=P(T_0^e)$. By~\eqref{eq:normalizedline},
\[
 R\bigl(X,Y+VX,P(X^e)\bigr)=0.
\]
At $X=T_0$ this gives $R(T_0,Y+VT_0,Z)=0$ and let us write 
$S(Y,V,Z)\coloneq R_Z(T_0,Y+VT_0,Z)$ for simplicity. Then since $(\alpha,\beta)\notin\mathcal E$ we have that
$R_Z(X,Y+VX,P(X^e))$ is nonzero over $\F_q$, so $S(Y,V,Z)\ne0$ when plugging in $T_0$.

We can thus apply Lemma~\ref{lem:hensel-lifting} to construct a formal
power series $E$ such that
\[
 P(X^e)-Z=E(X-T_0)=\sum_{j\ge1}c_j(Y,V,Z)(X-T_0)^j.
\]
Since $e$ is a power of the characteristic, $X^e-T_0^e=(X-T_0)^e$.
Thus $c_{ej}(Y,V,Z)$ is the coefficient of $(X-T_0^e)^j$ in $P(X)$
for $1\le j\le d$.

We now want to turn the equation above from Hensel lifting into a reconstruction map, which recovers $P$ from $\alpha$, $\beta$, and the initial value $P(T_0^e)$. Formally, we define
\[
 \Phi(Y,V,Z)\coloneq\left(Y^e,\;V^e,\;
 Z+\sum_{j=1}^{d}c_{ej}(Y,V,Z)(X-T_0^e)^j\right),
\]
where we think of the domain of the map as the set of triples $(Y,V,Z)$ for which Hensel lifting applies, or
\[
 \mathcal U\coloneq\{(Y,V,Z)\in\mathbb{K}^3:
 R(T_0,Y+VT_0,Z)=0,\quad S(Y,V,Z)\ne0\}.
\]
Then $\Phi$ gives a map from $\mathcal U$ to $\mathcal A$. When identifying the polynomial in the third coordinate with $\mathcal A$, we express it using its $d+1$ coefficients in the basis $1, X-T_0^e, \ldots, (X-T_0^e)^d$, so 
\[
 \Phi(Y,V,Z)
 =\Bigl(Y^e,V^e,Z,\underbrace{c_e(Y,V,Z),\ldots,c_{ed}(Y,V,Z)}_{j=1,\ldots,d}\Bigr)
 \in\mathbb{K}^2\times\mathbb{K}^{d+1}.
\]

Summarizing, we have now shown that each triple of interest
$(\alpha,\beta,P)\in\mathcal R_i$, can be obtained through
\[
 \Phi\bigl(\alpha^{1/e},\beta^{1/e},P(T_0^e)\bigr)=(\alpha,\beta,P),
\]
with $(\alpha^{1/e},\beta^{1/e},P(T_0^e))\in\mathcal U$. Hence, $\mathcal R_i\subseteq\Phi(\mathcal U)$.

To conclude, we bound the dimension and degree $\overline{\mc{U}}$, show that the map $\Phi$ is a rational map with numerators and common denominators of bounded degree, and then apply Lemma~\ref{lem:rational-image}.

First, by Lemma~\ref{lem:hypersurface}, $\dim\overline{\mathcal U}\le2$ and $\deg\overline{\mathcal U}\le h$. Second, for each $1\le j\le d$, Lemma~\ref{lem:hensel-lifting} gives
\[
 c_{ej}(Y,V,Z)=\frac{N_{ej}(Y,V,Z)}{S(Y,V,Z)^{2ej-1}},\qquad
 \deg N_{ej}\le1+(2ej-1)(h-1).
\]
We write every coordinate of
$\Phi$ with common denominator $S^{2ed-1}$. This yields, 
\[
 \frac{1}{S^{2ed-1}}\left(
 S^{2ed-1}Y^e,\ S^{2ed-1}V^e,\ S^{2ed-1}Z,\
 \bigl(S^{2e(d-j)}N_{ej}\bigr)_{j=1}^d
 \right),
\]
for the last $d$ coordinates of $\Phi$. Since $\deg S\le h-1$, the denominator has degree at most
$(2ed-1)(h-1)$. The numerator degrees are at most $e+(2ed-1)(h-1)$
for the first two coordinates and $1+(2ed-1)(h-1)$ for the polynomial
coefficients.

Finally, we let
\[
 V_i\coloneq\overline{\Phi(\mathcal U)}.
\]
Applying Lemma~\ref{lem:rational-image} to $\overline{\mathcal U}$ with
numerator and denominator degrees at most $2edh$ gives
\[
 \dim V_i\le2,\qquad \deg V_i\le h(2edh)^2=4d^2e^2h^3,
\]
and $\mathcal R_i\subseteq V_i$ for each $i \in [r]$. Every triple in the claim belongs to some $\mathcal R_i$ by
Lemma~\ref{lem:excluded}, so the $V_i$'s give a covering of all triples in the claim.
\end{proof}

\begin{proof}[Proof of Lemma~\ref{lem:nonpersistent}]
We prove the same bound for all nonpersistent triples
$(\alpha,\beta,P)$ with $(\alpha,\beta)\notin\mathcal E$,
$\deg P\le d$, and $|S_{\alpha,\beta}(P)|\ge A$.
The same bound holds for $|\mathcal N|$, since each of its
coefficient pairs occurs in at least one such triple.

By Lemma~\ref{lem:roots}, every triple being counted satisfies
\eqref{eq:lineidentity} and, by Claim~\ref{lem:envelope}, belongs to
$\bigcup_i V_i$. Apply Claim~\ref{lem:agreement-incidence} to each
$V_i$ with the hyperplanes $\mathcal H_z$. Summing over factors,
using $e_i\le M$ and $\sum_i h_i^3\le D^3$ from~\eqref{eq:factor-bounds},
we bound the number of
nonpersistent triples by
\[
 4d^2v^2\sum_i e_i^2h_i^3\le4d^2M^2D^3v^2.
\]
\end{proof}
Having bounded $|\mc{E}|$, $|\mc{P}|$, and $|\mc{N}|$, we can now prove Lemma~\ref{lem:counting}.
\begin{proof}[Proof of Lemma~\ref{lem:counting}]
The parameter definitions give
\[
 D\le\frac{9}{2\rho\eta^2},\qquad
 dM\le\frac{3n}{\eta},\qquad v\le\frac2\eta.
\]
Also, by~\eqref{eq:parameters},
\[
 \frac{B}{n}-\sqrt{\frac{d+1}{n}}
 \ge\frac\eta2-\frac1n-\frac1{2n\sqrt\rho}
 \ge\frac\eta4,
\]
so
\[
 \ell\le\frac{1}{B/n-\sqrt{(d+1)/n}}\le\frac4\eta.
\]

Every coefficient pair counted in the lemma belongs to
$\mathcal E\cup\mathcal P\cup\mathcal N$.
Substituting these estimates into the bounds from
\Cref{lem:excluded,lem:persistent,lem:nonpersistent}, and using
$\max\{q,\binom n2\}\le q+n^2/2$, bounds their total number by
\[
\begin{aligned}
 |\mathcal E|+|\mathcal P|+|\mathcal N|
 &\le D^2+\ell(q+n^2/2)+4(dM)^2D^3v^2\\
 &\le\frac{4q}{\eta}+\frac{2^{14}n^2}{\rho^3\eta^{10}}.
\end{aligned}
\]
\end{proof}

Theorem~\ref{thm:main} gives a simplified bound when
$n\eta^2\ge4/\rho$. This condition may fail at low rates. In that
case, one can choose $m,D$ directly and use
the finite counts in the proof of Lemma~\ref{lem:counting}.

\begin{theorem}\label{thm:finite-mca}
Set $0<\eta<1-\sqrt\rho$ and for positive integers $m,D$, set $M$ by~\eqref{eq:param}, and
suppose that $M\ge1$, the interpolation inequality~\eqref{eq:dimension}
holds, and $B^2>n(d+1)$. Then the MCA error with one-coordinate
agreement loss satisfies
\begin{equation}\label{eq:finite-mca-error}
\begin{aligned}
 \Err_{\rm aff}
 &\le\frac{\displaystyle
 \left\lfloor\frac{n(B-d-1)}{B^2-n(d+1)}\right\rfloor
 \max\{q,\tbinom n2\}}{q^2}\\
 &\quad+\frac{D^2+4d^2M^2D^3\left(\frac{n}{A-B}\right)^2}{q^2}.
\end{aligned}
\end{equation}
\end{theorem}
\begin{proof}
The interpolation inequality ensures that the construction in
Lemma~\ref{lem:roots} produces a nonzero explainer for these choices
of $m,D$. The condition $B^2>n(d+1)$ allows the Johnson bound in
Lemma~\ref{lem:persistent} to be applied. The remaining root-counting
arguments use the degree bounds on the explainer. Summing the bounds from
\Cref{lem:excluded,lem:persistent,lem:nonpersistent} gives
\eqref{eq:finite-mca-error}.
\end{proof}

When $q\ge\binom n2$, the first term is simply the rounded Johnson
bound divided by $q$.
For example, with $n=2^{15}$, $\rho=1/2048$, $\eta=0.05$, and
$q=2^{128}$, we have $d=16$, $A=2363$, and $B=1543$.
Choosing $m=1$ and $D=M=28$ satisfies the interpolation inequality
and gives
\[
 \Err_{\rm aff}\le27\cdot2^{-128}+2^{-211}
 \approx2^{-123.25},
\]
even though the condition of Theorem~\ref{thm:main} fails.

\section{Adjustments for Equal-degree, linear constraints, and interleaving}\label{sec:equal-degree}
For applications to proof systems, we need to modify the affine combiner to
handle a few additional requirements, namely equal degree bounds, linear
constraints, and higher-arity interleavings. We give adaptations that attain these properties without substantially compromising the MCA error from \cref{thm:main}.

\paragraph{Equal degree bounds.}
In most IOPPs, it is convenient for the two received words in a fold to have the same degree bound. To handle this, we can modify the combiner so that both words have affine coefficients. Sample independent uniform $\theta,\alpha,\beta\in\F_q$ and define
\begin{equation}\label{eq:equal-degree-combiner}
\begin{aligned}
 w_{\theta,\alpha,\beta}(z)
 &\coloneq g(z)+(\alpha+\beta z)(f(z)+\theta g(z))\\
 &=(\alpha+\beta z)f(z)+(1+\theta\alpha+\theta\beta z)g(z).
\end{aligned}
\end{equation}
For received words $f,g:\calD\to\F_q$, define
\[
 S_{\theta,\alpha,\beta}(P)
 \coloneq\{x\in\calD:P(x)=w_{\theta,\alpha,\beta}(x)\}.
\]
For a fixed choice of $(\theta,\alpha,\beta)$, say polynomials $F,G\in\F_q[X]$
are admissible witnesses for $P$ with agreement loss at most $s$ if
\begin{equation}\label{eq:equal-degree-witnesses}
\begin{gathered}
 \deg F,\deg G\le d-1,\qquad
 P=G+(\alpha+\beta X)(F+\theta G),\\
 \bigl|\{x\in S_{\theta,\alpha,\beta}(P):
 F(x)\ne f(x)\text{ or }G(x)\ne g(x)\}\bigr|\le s.
\end{gathered}
\end{equation}
We say that the equal-degree combiner has MCA error at most
$\varepsilon$ with agreement loss at most $s$ on $\calD$ if, for every
pair of received words $f,g:\calD\to\F_q$,
\[
 \Pr_{\theta,\alpha,\beta\,\leftarrow\,\F_q}
 \left[\begin{array}{l}
 \exists P\in\F_q[X],\ \deg P\le d:\\
 |S_{\theta,\alpha,\beta}(P)|\ge A\text{ and }P\text{ has no}\\
 \text{admissible witnesses with agreement loss at most }s
 \end{array}\right]\le\varepsilon.
\]
Similar to our notation for MCA error for affine combiners, we denote the smallest such $\varepsilon$ by
$\Err_{\rm eq}(n,\rho,A,s,q)$.

\paragraph{Linear constraints.}
In a typical proof system, the prover sends, along with the received words $f,g$, linear claims of the form
\[
 \langle m_f,v\rangle=\gamma_0,\qquad
 \langle m_g,v\rangle=\gamma_1,
\]
where $m_f,m_g$ are the messages that underlie the received
words $f,g$, and $v\in\F_q^{d+1}$ is a fixed public weight vector. In the case of Reed-Solomon codes, $m_f, m_g$ might be the $d$ coefficients of the degree $<d$ polynomial equal to $f, g$ respectively. That is, suppose that for honest degree $<d$ polynomials $F,G$, we have $m_f=\operatorname{coef}(F)$ and
$m_g=\operatorname{coef}(G)$, where $\operatorname{coef}(U)$ denotes
the coefficient vector of $U$ in the monomial basis
$(1,X,\ldots,X^d)$.

We want to carry these claims to a linear constraint on the coefficients of the
\[
 P=G+\alpha(F+\theta G)+\beta X(F+\theta G).
\]
The first two terms contribute
$\gamma_1+\alpha(\gamma_0+\theta\gamma_1)$ to its inner product with
$v$. The remaining term requires a claim on the message of
$X(F+\theta G)$, which the two original claims do not determine.
After receiving $\theta$, the prover therefore sends one field element
$c$, intended to equal
\[
 \bigl\langle\operatorname{coef}(X(F+\theta G)),v\bigr\rangle.
\]
Only then are $\alpha,\beta$ sampled. Writing
$m_P=\operatorname{coef}(P)$, the new claim on $m_P$ is
\begin{equation}\label{eq:equal-degree-linear-claim}
 \langle m_P,v\rangle
 =\gamma_1+\alpha(\gamma_0+\theta\gamma_1)+\beta c.
\end{equation}
For honest messages, the claim holds by linearity.

We show that our equal degree combiners can handle linear claims in the sense above with small additional proximity loss. 
\begin{theorem}\label{thm:equal-degree}
For the combiner~\eqref{eq:equal-degree-combiner}, the MCA error with
agreement loss $s=1$ satisfies
\begin{equation}\label{eq:equal-degree-mca-error}
\begin{aligned}
 \Err_{\rm eq}\Bigl(
 \underbrace{n}_{\substack{\text{block}\\\text{length}}},
 \underbrace{\rho}_{\substack{\text{degree}\\\text{fraction}}},
 \underbrace{A}_{\substack{\text{agreement}\\\text{threshold}}},
 \underbrace{1}_{\substack{\text{agreement}\\\text{loss}}},
 \underbrace{q}_{\substack{\text{field}\\\text{size}}}
 \Bigr)
 &\le\Err_{\rm aff}+\frac{\ell_0}{q}\\
 &\le\frac{4/\eta+\ell_0}{q}
       +\frac{2^{14}n^2}{\rho^3\eta^{10}q^2},
\end{aligned}
\end{equation}
where $\ell_0=\floor{n(A-1-d)/((A-1)^2-nd)}$.

Moreover, for the combination with linear claims as described above, the MCA
guarantee implies that, except with probability at most
\begin{equation}\label{eq:equal-degree-error}
 \Err_{\rm aff}+\frac{2\ell_0}{q}
 \le\frac{4/\eta+2\ell_0}{q}
       +\frac{2^{14}n^2}{\rho^3\eta^{10}q^2},
\end{equation}
every candidate $P$ of degree at most $d$ with
$|S_{\theta,\alpha,\beta}(P)|\ge A$ satisfying
\eqref{eq:equal-degree-linear-claim} has admissible witnesses $F,G$
with agreement loss at most one satisfying
\begin{equation}\label{eq:equal-degree-claim-witnesses}
\begin{gathered}
 \langle\operatorname{coef}(F),v\rangle=\gamma_0,\qquad
 \langle\operatorname{coef}(G),v\rangle=\gamma_1,\\
 \langle\operatorname{coef}(X(F+\theta G)),v\rangle=c.
\end{gathered}
\end{equation}
This holds for every prover strategy.
\end{theorem}
\begin{proof}
Consider the list
\[
 \mathcal L\coloneq\left\{(F,G)\in\F_q[X]_{\le d}^2:
 \bigl|\{z\in\calD:F(z)=f(z),\ G(z)=g(z)\}\bigr|\ge A-1\right\}.
\]
By~\eqref{eq:parameters}, we have $n\eta>1$, so
$A-1>\sqrt{nd}$ and the denominator defining $\ell_0$ is positive.
Two distinct pairs have at most $d$ common agreement coordinates,
since at least one of their polynomial differences is nonzero and
has degree at most $d$. Lemma~\ref{lem:johnson}, with
$a=(A-1)/n$ and $b=d/n$, gives $|\mathcal L|\le\ell_0$.
This list is independent of all challenges.

For each fixed $\theta$, apply Theorem~\ref{thm:main} to the received
words $(g,f+\theta g)$. Except with probability $\Err_{\rm aff}$ over
$(\alpha,\beta)$, every candidate under consideration has witnesses
$U,V$ with $\deg U\le d$, $\deg V\le d-1$, and
$P=U+(\alpha+\beta X)V$, explaining all but one of its agreements.
Set $G=U$ and $F=V-\theta U$. Then $(F,G)\in\mathcal L$,
the combination identity and agreement guarantee
in~\eqref{eq:equal-degree-witnesses} hold, and
\[
 [X^d]F+\theta[X^d]G=0.
\]
For each pair in $\mathcal L$ with $\deg G=d$, this equation
determines exactly one $\theta$. With probability at most
$\ell_0/q$ over $\theta$, any such pair can occur.
Outside these choices, $\deg G\le d-1$ and
$F=V-\theta G$ also has degree at most $d-1$.
This proves the first assertion.

For the claim guarantee, condition on $\theta$ and the prover's
message $c$. Fix a pair $(F,G)\in\mathcal L$ with both degrees at
most $d-1$, and define
\[
\begin{gathered}
 \Delta_F=\langle\operatorname{coef}(F),v\rangle-\gamma_0,\qquad
 \Delta_G=\langle\operatorname{coef}(G),v\rangle-\gamma_1,\\
 \Delta_X=\langle\operatorname{coef}(X(F+\theta G)),v\rangle-c.
\end{gathered}
\]
For its combined polynomial, the residual
in~\eqref{eq:equal-degree-linear-claim} is
\[
 \Delta_G+\alpha(\Delta_F+\theta\Delta_G)+\beta\Delta_X.
\]
If any equality in~\eqref{eq:equal-degree-claim-witnesses} fails,
this is a nonzero affine polynomial in $\alpha,\beta$ and vanishes
with probability at most $1/q$. A union bound over the at most
$\ell_0$ pairs, followed by the two exceptional events above,
proves~\eqref{eq:equal-degree-error}.
\end{proof}

Since $\ell_0=O_{\rho,\eta}(1)$, both guarantees retain error
$O_{\rho,\eta}(1/q+n^2/q^2)$.

\paragraph{Round complexity.}
The schedules above introduce one extra interaction round per fold.
This is done mainly for simplicity. An optimization removes this round
at the cost of one additional
field element per functional, while preserving the soundness bounds
for the original claims.

The extra round arises because the prover's
auxiliary claims are sent after $\theta$ and before the remaining
challenges. For a single functional, the prover can instead announce
claims $c_F=\langle\operatorname{coef}(XF),v\rangle$
and $c_G=\langle\operatorname{coef}(XG),v\rangle$ alongside
$\gamma_0,\gamma_1$, before any challenges are sampled. The verifier then samples $\theta,\alpha,\beta$ together and
uses $c=c_F+\theta c_G$ in~\eqref{eq:equal-degree-linear-claim}.
The bound in~\eqref{eq:equal-degree-error} still
protects the original claims
$\langle\operatorname{coef}(F),v\rangle=\gamma_0$ and
$\langle\operatorname{coef}(G),v\rangle=\gamma_1$:
conditioned on $\theta$, the derived $c$ is fixed before
$\alpha,\beta$. This guarantee need not certify $c_F,c_G$
individually.

\paragraph{Higher-arity interleavings.}
We next fold many received words using shared challenges, requiring
agreement on a domain shared by all words. Start with received words
$f_1,\ldots,f_k:\calD\to\F_q$, where $k\ge2$ is a power of two, and
set $r\coloneq\log_2 k$. Sample independent uniform triples
$(\theta_j,\alpha_j,\beta_j)\in\F_q^3$ for $1\le j\le r$.
Starting with $f_i^{(0)}=f_i$, define round $j$ by
\[
 f_i^{(j)}(z)
 =f_{2i}^{(j-1)}(z)+(\alpha_j+\beta_j z)
   \bigl(f_{2i-1}^{(j-1)}(z)+\theta_j f_{2i}^{(j-1)}(z)\bigr),
 \qquad 1\le i\le k/2^j.
\]
Thus each round halves the number of rows, with the same challenges
for every pair. Denote the final word by $w\coloneq f_1^{(r)}$.
Assume Theorem~\ref{thm:equal-degree} applies in round $j$ with degree
bound $d+j$ and agreement threshold $A$, and let $\varepsilon$ be the
maximum of its MCA bounds~\eqref{eq:equal-degree-mca-error} over the
$r$ rounds.

\begin{theorem}\label{thm:interleaved}
Except with probability at most $r\varepsilon$, every polynomial $P$
of degree at most $d+r$ with an agreement domain $S$ of size at least
$A+r$ with $w$ has polynomial witnesses $F_1,\ldots,F_k$ of degree
at most $d$ whose folding with the same challenges gives $P$ and for which
\[
 \bigl|\{z\in S:\exists i,\ F_i(z)\ne f_i(z)\}\bigr|\le r.
\]
\end{theorem}
\begin{proof}
We first prove the one-round guarantee for $t$ simultaneous
combinations. Fix received words $f_i,g_i:\calD\to\F_q$ for
$1\le i\le t$, and use the same uniform $\alpha,\beta\in\F_q$ in
all combinations $f_i(z)+(\alpha+\beta z)g_i(z)$.

For a tuple $P=(P_1,\ldots,P_t)$, define its shared agreement domain by
\[
 S_{\alpha,\beta}(P)\coloneq\{z\in\calD:
 P_i(z)=f_i(z)+(\alpha+\beta z)g_i(z)\text{ for every }i\}.
\]
Tuples $F=(F_1,\ldots,F_t)$ and $G=(G_1,\ldots,G_t)$ of polynomials
over $\F_q$ are admissible witnesses for $P$ if
\[
\begin{gathered}
 \deg F_i\le d,\quad \deg G_i\le d-1,\quad
 P_i=F_i+(\alpha+\beta X)G_i\qquad(1\le i\le t),\\
 \bigl|\{z\in S_{\alpha,\beta}(P):
 \exists i,\ F_i(z)\ne f_i(z)\text{ or }G_i(z)\ne g_i(z)\}\bigr|\le1.
\end{gathered}
\]
Thus the agreement loss of one coordinate is shared by all rows.

We establish the one-round bound
\begin{equation}\label{eq:interleaved-mca-error}
 \Pr_{\alpha,\beta\,\leftarrow\,\F_q}
 \left[\begin{array}{l}
 \exists P\in(\F_q[X]_{\le d})^t:\\
 |S_{\alpha,\beta}(P)|\ge A\text{ and }P\text{ has no}\\
 \text{admissible witness tuples}
 \end{array}\right]
 \le\frac{4}{\eta q}+\frac{2^{14}n^2}{\rho^3\eta^{10}q^2}.
\end{equation}
Fix an $\F_q$-basis
$\zeta_1,\ldots,\zeta_t$ of $\F_{q^t}$. Pack each tuple into one $\F_{q^t}$-valued
word or polynomial by taking its linear combination with coefficients
$\zeta_i$. Since $\calD\subseteq\F_q$ and the challenges lie in
$\F_q$, packing commutes with evaluation and with the combiner.
Equality of packed symbols is equivalent to equality in every row.

We show that the proof of Theorem~\ref{thm:main} gives the same bound
for the packed words, with polynomial coefficients in $\F_{q^t}$ and
challenges in $\F_q$. Interpolation, factorization, simple-root
lifting, and the algebraic degree bounds hold over $\F_{q^t}$ with the same
parameters. The excluded coefficient pairs are still counted by line
factors; restricting the challenges to $\F_q^2$ cannot increase their
number. Taking $e_i$-th roots also preserves $\F_q$.

For persistent triples, the Johnson bound counts shared agreement
sets and is unchanged by extending the coefficient field. A rational
pair over $\F_{q^t}$ that fails the polynomial degree bounds imposes a
nonzero affine equation on $\alpha,\beta$. Even with coefficients in
$\F_{q^t}$, such an equation has at most $q$ solutions in $\F_q^2$: fixing
one coordinate determines the other whenever its coefficient is nonzero.
For a pair satisfying the degree bounds, two accidental agreement
coordinates determine at most one challenge, giving the same
$\binom n2$ bound.

The nonpersistent bound in Claim~\ref{lem:agreement-incidence} counts
all $\F_{q^t}$-rational triples in the algebraic cover, so it also bounds the
subset whose coefficient pairs lie in $\F_q^2$.
In its case $c>B$, the affine linear system defining $W$ is now over
$\F_{q^t}$. If it contains a triple and has positive dimension, it contains
a second $\F_{q^t}$-rational triple; the same subtraction gives a persistent
rational explanation over $\F_{q^t}$. The second triple need not have its
coefficient pair in $\F_q^2$. The hyperplane-section induction and
the reconstruction cover therefore retain their bounds.

Consequently, the number of exceptional base-field challenges is at most
\[
 D^2+\ell\max\{q,\tbinom n2\}+4d^2M^2D^3v^2.
\]
The estimates in the proof of Lemma~\ref{lem:counting}, followed by
division by $q^2$, give~\eqref{eq:interleaved-mca-error}.
Unpacking the witness polynomials gives the required tuples, with
at most one discarded coordinate shared by all rows.

For the equal-degree and linear-claim conclusions, apply the same
packing to the witness lists, claimed values, and auxiliary messages.
Distinct packed polynomial pairs have at most $d$ common agreement
coordinates, so their list has size at most $\ell_0$. The degree test
in $\theta$ has at most one base-field solution for each pair that
fails the input degree bounds. The affine zero counts in the
claim-binding arguments are unchanged: each nonzero equation over $\F_{q^t}$
has at most $q$ base-field pairs. Common weight vectors make the
inner products commute with packing. Thus the proof of
Theorem~\ref{thm:equal-degree} gives its stated bounds without a
factor of $t$.

Now apply the interleaved equal-degree bound in each of the $r$ rounds,
starting with $t=k/2$ pairs. Conditioning on the preceding rounds
fixes the current received words before the fresh challenges are sampled,
so each round fails with probability at most $\varepsilon$. A union
bound gives total failure probability at most $r\varepsilon$.
Outside these failures, work backward from $P$. Each round recovers
polynomials for all parent rows and discards at most one shared
agreement coordinate. Since $|S|\ge A+r$, every intermediate tuple
retains at least $A$ agreements, so the one-round guarantee continues
to apply. After $r$ steps, the recovered polynomials have degree at
most $d$ and agree with all original words on all but at most $r$
coordinates of $S$. Their combination identities compose to give $P$.
\end{proof}

\begin{remark}\label{rem:flock-folding}
Flock uses Ligerito~\cite{Ligerito} to encode each message row with the
same Reed--Solomon code and fold the encoded rows by multilinear partial
evaluation~\cite[Appendices~C.1--C.2]{Flock}. Each scalar folding round
replaces every pair $u,v$ by $(1-\lambda)u+\lambda v$, using the same
fresh uniform $\lambda$ for all pairs. Theorem~\ref{thm:interleaved}
gives the MCA guarantee for an affine variant of this row folding,
with common-weight linear constraints handled as above.

We now spell out the MCA error and the grinding needed to reach a
$128$-bit target for this contribution. When $q\ge\binom n2$,
\eqref{eq:finite-mca-error} gives
\[
 \Err_{\rm aff}\le\frac{\ell}{q}+\delta,
 \qquad
 \delta\coloneq\frac{D^2+4d^2M^2D^3\bigl(n/(A-B)\bigr)^2}{q^2}.
\]

For $n=2^{20}$, $d=2^{18}$, $\eta=0.05$, $A=576717$, and
$q=2^{128}$, the choices $m=3$, $D=49$, $M=6$, and $B=567980$
satisfy the interpolation dimension inequality and give
$\ell\le6$, $\ell_0\le5$, and $\delta<2^{-182}$.
Thus the original affine combiner has MCA error at most
$6\cdot2^{-128}+\delta$, or approximately $125.42$ bits of error
($-\log_2$ of the bound). For equal degree bounds,
\eqref{eq:equal-degree-mca-error} adds at most $5\cdot2^{-128}$,
giving error at most $11\cdot2^{-128}+\delta$, or about $124.54$ bits.
Carrying a common-weight linear claim adds another
$5\cdot2^{-128}$ by~\eqref{eq:equal-degree-error}, giving error at
most $16\cdot2^{-128}+\delta$, or about $124$ bits.

In the standard Fiat--Shamir grinding accounting, $b$ bits of proof
of work reduce the protected round's error contribution by a factor
of $2^b$~\cite[Section~3.2]{BT24}. Consequently, three bits of grinding
bring the original combiner's MCA contribution below $2^{-128}$,
and four bits suffice for the equal-degree version. With the linear
claim, four bits cover the leading term $16\cdot2^{-128}$; five bits
also absorb the positive correction and give a bound below
$2^{-128}$. 
\end{remark}

\section{Separation from scalar combiners and Necessity of Agreement Loss}\label{sec:separation}
In this section we show a separation between affine and scalar combiners by adapting the counterexample of~\cite[Remark~1.1]{BCIKS}. In particular, in the regime $n \ll q$, \cref{thm:main} shows that affine combiners can achieve error $O(1/q)$, while the separation result below shows that scalar combiners must have error $\Omega(n/q)$, even for larger agreement losses $s > 1$. For scalar combiners, write
\[
 S_\alpha(P)\coloneq\{z\in\calD:P(z)=f(z)+\alpha g(z)\}.
\]
and let
$\Err_{\rm sc}(n,d,A,s,q)$ denote the MCA error at agreement threshold
$A$, defined as for affine combiners but with a uniform scalar multiplier
$\alpha$ instead and degree bounds $\deg F,\deg G\le d$.

\begin{proposition}\label{prop:scfloor}
Let $s\ge0$ be an integer and suppose $A-s-1>d$.
There are received words $f,g$ for which no polynomial of degree at
most $d$ agrees with $g$ on $A-s$ coordinates,
while
\begin{equation}\label{eq:scalar-necessary}
 \Err_{\rm sc}(n,d,A,s,q)
 \ge\Pr_{\alpha\leftarrow\F_q}\bigl[|S_\alpha(0)|=A\bigr]
 \ge\frac{1+\lfloor(n-A)/(s+1)\rfloor}{q}.
\end{equation}
\end{proposition}
\begin{proof}
Choose $C\subseteq\calD$ of size $A-s-1$ and define
\[
 V_C(X)\coloneq\prod_{z'\in C}(X-z'),\qquad g(z)\coloneq V_C(z).
\]
For every polynomial $G$ of degree at most $d$, the difference
$V_C-G$ has degree $A-s-1$. Thus $G$ agrees with $g$ on at most
$A-s-1$ coordinates.

Partition $\calD\setminus C$ into blocks of size $s+1$ and, if needed,
one smaller nonempty block. Assign a distinct $\gamma_T\in\F_q$ to each block
$T$; this is possible since there are at most $n\le q$ blocks.
Define
\[
 f(z)\coloneq0\quad(z\in C),\qquad
 f(z)\coloneq-\gamma_Tg(z)\quad(z\in T).
\]
Because $g$ is nonzero outside $C$, each block $T$ of size $s+1$
satisfies $S_{\gamma_T}(0)=C\cup T$, of size $A$.
With agreement loss $s$, at least $A-s$ agreements must be preserved,
and no degree-$d$ polynomial can agree with $g$ on that many
coordinates. Hence $P=0$ has no admissible witnesses
for each of these distinct challenges. There are
$1+\lfloor(n-A)/(s+1)\rfloor$ blocks of size $s+1$.
Since $\alpha$ is uniform in $\F_q$, we obtain~\eqref{eq:scalar-necessary}.
\end{proof}

For the comparison with affine combiners, set $s=1$ and write
$\Err_{\rm sc}\coloneq\Err_{\rm sc}(n,d,A,1,q)$.
For fixed $\rho,\eta$ and sufficiently large $n$, we have $A-2>d$
and $n-A=\Theta_{\rho,\eta}(n)$. By Proposition~\ref{prop:scfloor} and
Theorem~\ref{thm:main},
\[
 \Err_{\rm sc}=\Omega_{\rho,\eta}(n/q),\qquad
 \Err_{\rm aff}=O_{\rho,\eta}(1/q)
 \quad\text{when }q\ge n^2.
\]
Thus the affine combiner improves the worst-case MCA error by a factor of
$\Omega_{\rho,\eta}(n)$ at the same agreement threshold and with the
same agreement loss $s=1$.

We also show that without allowing the one point of agreement loss, affine combiners cannot achieve the same low error as with $s=1$.

\begin{proposition}\label{prop:affine-zero-loss}
Suppose $A-1>d$. Then
\begin{equation}\label{eq:lower}
 \Err_{\rm aff}(n,\rho,A,0,q)
 \ge\frac{(n-A+1)(q-1)}{q^2}
 \ge\frac{n-A+1}{2q}.
\end{equation}
\end{proposition}
\begin{proof}
Choose $C\subseteq\calD$ of size $A-1$. Let $f\coloneq0$, and let
$g$ be zero on $C$ and one elsewhere. For each $z\in\calD\setminus C$,
there are $q-1$ nonzero pairs $(\alpha,\beta)$ with $\alpha+\beta z=0$.
Every such pair has $\beta\ne0$ and
\[
 S_{\alpha,\beta}(0)=C\cup\{z\},\qquad |S_{\alpha,\beta}(0)|=A.
\]
Any polynomial $G$ of degree at most $d-1$ agreeing with $g$ on $C$
must be zero, since $|C|=A-1>d$. It therefore disagrees with $g$ at
$z$, so $P=0$ has no witnesses with agreement loss $s=0$.

The nonzero pairs counted for distinct $z$ are disjoint. Dividing
$(n-A+1)(q-1)$ by $q^2$ proves the first inequality
in~\eqref{eq:lower}; $q-1\ge q/2$ proves the second.
\end{proof}

\paragraph{Constant fractional agreement loss.}
Even when a constant fraction of agreement loss is allowed, we show
that scalar combiners cannot achieve $O(1/q)$ MCA error even on
additive evaluation domains. We adapt the DEEP-FRI subspace
counterexample~\cite[Appendix~B]{BGKS20} to degree bound $d=n/4$
by multiplying both received words and their candidate polynomials
by a common vanishing polynomial.

\begin{proposition}\label{prop:scalar-constant-loss}
Suppose $\F_q$ has characteristic two, $n\ge16$ is a power of two,
$q\ge n$, $d=n/4$, and $A\le9n/16$.
For every $\F_2$-linear evaluation domain $\calD\subseteq\F_q$ of size $n$
and every integer $0\le s<3n/16$,
\begin{equation}\label{eq:scalar-constant-loss}
 \Err_{\rm sc}(n,d,A,s,q)\ge\frac{n-8}{q}.
\end{equation}
\end{proposition}
\begin{proof}
We construct $n-8$ combinations with $9n/16$ agreements, while every
degree-at-most-$d$ polynomial agrees with $g$ on at most $3n/8$
coordinates. The gap of $3n/16$ makes these combinations exceptional
for every allowed loss $s$ in the statement.

Choose a codimension-three subspace $\mathcal U\subseteq\calD$,
so $|\mathcal U|=n/8$, and define
\[
 W(X)\coloneq\prod_{\gamma\in\mathcal U}(X-\gamma).
\]
For $z\in\calD$, set
\[
 f(z)\coloneq W(z)z^{n/2},\qquad
 g(z)\coloneq W(z)z^{n/4}.
\]
For every polynomial $G$ of degree at most $d$,
the polynomial $W(X)X^{n/4}-G(X)$ is nonzero and has degree $3n/8$.
Thus $G$ agrees with $g$ on at most $3n/8$ coordinates.

For each $\F_2$-linear hyperplane $\mathcal H\subseteq\calD$, consider its
subspace polynomial
\[
 L_{\mathcal H}(X)\coloneq\prod_{\gamma\in\mathcal H}(X-\gamma).
\]
Such polynomials have only power-of-two exponents. This follows
inductively from $L_{\{0\}}(X)=X$: for an $\F_2$-subspace
$\mathcal W$ and $\gamma\notin\mathcal W$, we have
\[
 L_{\mathcal W+\F_2\gamma}(X)
 =L_{\mathcal W}(X)^2-L_{\mathcal W}(\gamma)L_{\mathcal W}(X).
\]
Consequently,
\[
 L_{\mathcal H}(X)
 =X^{n/2}+\alpha_{\mathcal H}X^{n/4}+R_{\mathcal H}(X),
 \qquad \deg R_{\mathcal H}\le n/8.
\]
Distinct hyperplanes give distinct coefficients $\alpha_{\mathcal H}$:
otherwise their subspace polynomials would differ by a polynomial
of degree at most $n/8$ vanishing on their intersection of size $n/4$.
They would therefore be equal, contradicting the distinct root sets.

Set $P_{\mathcal H}\coloneq-WR_{\mathcal H}$, so
$\deg P_{\mathcal H}\le n/4=d$. For $z\in\calD$,
\[
 f(z)+\alpha_{\mathcal H}g(z)-P_{\mathcal H}(z)
 =W(z)L_{\mathcal H}(z),
\]
and hence
\[
 S_{\alpha_{\mathcal H}}(P_{\mathcal H})
 =\mathcal U\cup\mathcal H.
\]
If $\mathcal H$ does not contain $\mathcal U$, then
$|\mathcal H\cap\mathcal U|=n/16$, giving
\[
 |S_{\alpha_{\mathcal H}}(P_{\mathcal H})|
 =\frac n8+\frac n2-\frac n{16}=\frac{9n}{16}\ge A.
\]
Of the $n-1$ hyperplanes of $\calD$, exactly $2^3-1=7$ contain
$\mathcal U$, leaving $n-8$ distinct challenges with this agreement.
For each of them, witnesses with agreement loss at most $s$ would
have to agree with $g$ on at least $9n/16-s>3n/8$ coordinates,
which is impossible. Dividing by $q$ proves
\eqref{eq:scalar-constant-loss}.
\end{proof}

Taking $s=\floor{\lambda n}$ for any fixed $0<\lambda<3/16$
gives scalar MCA error $\Omega(n/q)$, including for $1\%$ and $10\%$
agreement loss. At $\rho=1/4$, the condition on $A$ includes every
fixed Johnson slack $0<\eta\le1/16$. For sufficiently large $n$ and
$q\ge n^2$, Theorem~\ref{thm:main} gives affine MCA error
$O_{\rho,\eta}(1/q)$ already with one-coordinate loss, and hence also
with this larger allowed loss. Thus the separation persists when a
constant fraction of agreement may be discarded.

\ifrmext
\section{Extension to Reed--Muller codes}\label{sec:rm}

Fix an integer $k\ge1$ and a subfield $\F_{q_0}\subseteq\F_q$, and set
$n\coloneq q_0^k$, the size of the grid $\F_{q_0}^k\subseteq\F_q^k$. In this section,
degrees of polynomials in $\F_q[\mathbf X]=\F_q[X_1,\ldots,X_k]$ are
total degrees. Received words are functions on the grid, so their
coordinates are the grid points. For $\rho\in(0,1)$ such that
$d\coloneq\rho q_0$ is an integer, the Reed--Muller code consists of the
evaluations on $\F_{q_0}^k$ of polynomials of degree at most $d$, and it
has relative distance $1-\rho$. We call
$\rho$ the degree fraction; unlike for Reed--Solomon codes, it is not the
rate of the code when $k\ge2$.

\paragraph{Grid points on a hyperplane.}
Let $H=\{\mathbf x\in\F_q^k:\mathbf a\cdot\mathbf x=b\}$ be a hyperplane
of $\F_q^k$, where $\mathbf a\ne0$. The grid points in $H$, if any, form
a coset of the $\F_{q_0}$-linear subspace
$\{\mathbf z\in\F_{q_0}^k:\mathbf a\cdot\mathbf z=0\}$, which is proper
since $\mathbf a\ne0$. Therefore
\begin{equation}\label{eq:rm-hyperplane}
 |H\cap\F_{q_0}^k|\le n/q_0,
\end{equation}
with equality for the hyperplane $x_1=0$. Consequently, a set of more
than $n/q_0$ grid points is not contained in any hyperplane, so it
contains $k+1$ affinely independent points.

\paragraph{The affine combiner.}
For received words $f,g:\F_{q_0}^k\to\F_q$, we sample a uniform
$\boldsymbol\alpha=(\alpha_0,\ldots,\alpha_k)\in\F_q^{k+1}$ and form
\[
 w_{\boldsymbol\alpha}(\mathbf x)\coloneq
 f(\mathbf x)+L_{\boldsymbol\alpha}(\mathbf x)g(\mathbf x),
 \qquad
 L_{\boldsymbol\alpha}(\mathbf X)\coloneq\alpha_0+\alpha_1X_1+\cdots+\alpha_kX_k.
\]
As in Section~\ref{sec:our-results}, the factor $L_{\boldsymbol\alpha}$ is
the affine combiner, and for polynomials with
$\deg F\le d$ and $\deg G\le d-1$, the polynomial
$F+L_{\boldsymbol\alpha}G$ has degree at most $d$.

\paragraph{Definition of the MCA error.}
Let $1\le A\le n$ and $s\ge0$ be integers, and let $
 S_{\boldsymbol\alpha}(P)\coloneq
 \{\mathbf x\in\F_{q_0}^k:P(\mathbf x)=w_{\boldsymbol\alpha}(\mathbf x)\}$.
For $\boldsymbol\alpha$, polynomials $F,G\in\F_q[\mathbf X]$ of degrees at
most $d$ and $d-1$ respectively are admissible witnesses for
$P=F+L_{\boldsymbol\alpha}G$ with agreement loss at most $s$ if
\[
 \bigl|\{\mathbf x\in S_{\boldsymbol\alpha}(P):
 F(\mathbf x)\ne f(\mathbf x)\text{ or }G(\mathbf x)\ne g(\mathbf x)\}\bigr|\le s.
\]
As in Section~\ref{sec:our-results}, the MCA error with agreement loss
at most $s$ is the smallest $\varepsilon$ such that, for every pair
$f,g$, the probability over $\boldsymbol\alpha$ that some polynomial $P$
with $\deg P\le d$ and $|S_{\boldsymbol\alpha}(P)|\ge A$ has no
admissible witnesses is at most $\varepsilon$. We denote it by
$\Err_{\rm RM}(k,q_0,\rho,A,s,q)$.

\begin{theorem}\label{thm:rm}
Let $k\ge1$ be an integer and $\F_{q_0}$ a finite field, and set
$n\coloneq q_0^k$. Fix $\rho\in(0,1)$ with $\rho q_0$ an integer, let
$0<\eta<1-\sqrt\rho$, and set $A\coloneq\ceil{(\sqrt\rho+\eta)n}$.
Suppose $q_0\eta^2\ge4/\rho$. Then, for every finite field $\F_q$
containing $\F_{q_0}$, the MCA error with agreement loss $s=n/q_0$
satisfies
\begin{equation}\label{eq:rm-bound}
 \Err_{\rm RM}\Bigl(
 \underbrace{k}_{\substack{\text{number of}\\\text{variables}}},
 \underbrace{q_0}_{\substack{\text{subfield}\\\text{size}}},
 \underbrace{\rho}_{\substack{\text{degree}\\\text{fraction}}},
 \underbrace{A}_{\substack{\text{agreement}\\\text{threshold}}},
 \underbrace{n/q_0}_{\substack{\text{agreement}\\\text{loss}}},
 \underbrace{q}_{\substack{\text{field}\\\text{size}}}
 \Bigr)
 \le\frac{4}{\eta q^k}+\frac{2^{6k+8}n^{k+1}}{\rho^{k+2}\eta^{4k+6}q^{k+1}}.
\end{equation}
\end{theorem}

For $k=1$, Theorem~\ref{thm:rm} is Theorem~\ref{thm:main} for the
domain $\calD=\F_{q_0}$. For
fixed $\rho,\eta,k$ and $q\ge n^{k+1}$, the error is
$O_{\rho,\eta,k}(q^{-k})$.

The exponent $k$ cannot be improved. The $q$ challenges with
$\alpha_1=\cdots=\alpha_k=0$, a $q^{-k}$ fraction of all challenges, have
the constant multiplier $\alpha_0$, so a received word $g$ of degree $d$
leads to combined words of degree $d$; these have no admissible
witnesses, which require $\deg G\le d-1$. Indeed, take
$f=0$ and $g=X_1^d$. For each such challenge, the polynomial
$P=\alpha_0X_1^d$ agrees with $w_{\boldsymbol\alpha}$ on all $n$
coordinates. By the Schwartz--Zippel lemma, every polynomial of degree
at most $d-1$ disagrees with $X_1^d$ on at least
$(1-\rho)n=(q_0-d)n/q_0$ coordinates. This is more than $n/q_0$, since
the conditions of Theorem~\ref{thm:rm} imply $d\le q_0-2$: if
$d=q_0-1$, then $\eta<1-\sqrt\rho<1-\rho=1/q_0$, so
$q_0\eta^2<1/q_0<4/\rho$. Thus $P$ has no admissible witnesses, and
$\Err_{\rm RM}(k,q_0,\rho,A,n/q_0,q)\ge q/q^{k+1}=q^{-k}$.

\subsection{Proof idea}\label{sec:rm-idea}
We prove Theorem~\ref{thm:rm} in Appendix~\ref{app:rm-proof} by
following the proof of Theorem~\ref{thm:main}. The proof changes in
three places.

\paragraph{Agreement loss and the order of the error.}
Fix polynomials $F,G$ with $\deg F\le d$ and $\deg G\le d-1$, and call
an agreement of $P=F+L_{\boldsymbol\alpha}G$ with $w_{\boldsymbol\alpha}$
\emph{accidental} if it occurs at a coordinate $\mathbf z$ where $F,G$ do
not both agree with $f,g$. For each such coordinate, the value
$L_{\boldsymbol\alpha}(\mathbf z)$ is fixed as in the proof of
Lemma~\ref{lem:rational-pair}. In the univariate case, the challenge is
determined by two such values at distinct coordinates. Here, values at $k+1$ affinely
independent points are needed: values on a hyperplane
$\{L_{\boldsymbol\sigma}=0\}$ are the same for at least $q$ challenges,
since adding a multiple of $\boldsymbol\sigma$ to $\boldsymbol\alpha$
does not change them. A hyperplane contains at most $n/q_0$ grid points
by~\eqref{eq:rm-hyperplane}, so with agreement loss $n/q_0$, an
exceptional candidate has accidental agreements at $k+1$ affinely
independent points. Hence at most $\binom n{k+1}$ of the $q^{k+1}$
challenges are exceptional for the pair.

If $F,G$ are rational functions that do not satisfy these degree bounds,
we use, as in the proof of Lemma~\ref{lem:rational-pair}, that the challenges for which
$F+L_{\boldsymbol\alpha}G$ is a polynomial of degree at most $d$ form a
proper affine subspace. For $k\ge2$, a proper subspace can contain $q^k$
challenges, and the resulting error bound would only be of order $1/q$.
By Lemma~\ref{lem:rm-rational-pair}, this subspace has dimension
at most one: for any two of its elements
$\boldsymbol\alpha,\boldsymbol\alpha'$, the product
$L_{\boldsymbol\alpha-\boldsymbol\alpha'}G$ is a polynomial of degree at
most $d$, and $\boldsymbol\alpha-\boldsymbol\alpha'$ is therefore a
multiple of one fixed vector. Summing over the $O(1/\eta)$ persistent
pairs, with at most $q$ challenges per pair, we obtain a leading term
of order $1/(\eta q^k)$.

\paragraph{Reduction to univariate polynomials.}
The interpolation in Section~\ref{sec:interp} is univariate, so we first
map the Reed--Muller code into a Reed--Solomon code of length $n$ with
the same degree fraction~\cite{PW04}. This map identifies the grid with
the field $\F_n$ through an $\F_{q_0}$-linear bijection whose components,
polynomials of degree $n/q_0$, are substituted for $X_1,\ldots,X_k$.
Polynomials of degree at most $d$ become univariate
polynomials of degree at most $dn/q_0=\rho n$, so the agreement
threshold is the same as in Theorem~\ref{thm:main}. The affine
combiner, however, becomes a polynomial of degree $n/q_0$, so we
interpolate with the $(1,n/q_0,\rho n)$-weighted degree. Because of this
larger degree, Theorem~\ref{thm:rm} requires $q_0\eta^2\ge4/\rho$, while
Theorem~\ref{thm:main} for block length $n$ would require only
$n\eta^2\ge4/\rho$.

\paragraph{Nonpersistent candidates.}
The challenge now consists of $k+1$ field elements, so the nonpersistent
candidates lie in algebraic sets of dimension at most $k+1$, and the
agreement-incidence argument bounds their number by
$O_{\rho,\eta,k}(n^{k+1})$, so their contribution to the error is of
order $n^{k+1}/q^{k+1}$. The reconstructed univariate polynomials need
not have the form $\widetilde P$ for a Reed--Muller polynomial $P$,
so we intersect these algebraic sets with the linear subspace where they
do, which does not increase their dimension or degree. In the agreement-incidence argument,
the denominator $L_{\boldsymbol\sigma}$ of a persistent pair can vanish
at up to $n/q_0$ coordinates by~\eqref{eq:rm-hyperplane}, which is why
the persistence threshold is lowered by $n/q_0$. Simple-root lifting and the
agreement-incidence argument are otherwise unchanged, since they only
use that the multiplier is linear in $\boldsymbol\alpha$.

\fi

\section*{Acknowledgments}
\label{sec:acknowledgements}

We thank Ron Rothblum for his comments on an earlier version of this paper,
particularly on its connection to Flock. We also thank Ariel Gabizon for his
comments.

Eylon Yogev is supported by the Israel Science Foundation (Grant No.\ 2302/22) and the European Union (ERC, CRYPTOPROOF, 101164375). Views and opinions expressed are however those of the author(s) only and do not necessarily reflect those of the European Union or the European Research Council. Neither the European Union nor the granting authority can be held responsible for them.

\paragraph{AI usage.}
The authors were initially motivated to explore this direction by the
connection between proximity gaps and robust versions of the
line-versus-point test, as described in Section~\ref{sec:techniques}. In an earlier version of this paper, the authors were able to obtain proximity gaps for high-variate Reed-Muller codes over small fields up to their list-decoding radius using this agreement testing connection and the list-decoding results of \cite{GKZ08, BL18}. However, this result used combinatorial techniques and we were unable to obtain interesting results for Reed-Solomon codes using them.

The authors then asked ChatGPT 6 Astra to prove an improved proximity-gap result
for these combiners and immediately obtained \cref{thm:main}. The analysis, including all proofs appearing in
this paper, was carried out entirely with AI. The authors edited, verified, reorganized, and Lean-formalized the initial writeup. The Lean formalization was developed using ChatGPT 6 Astra, and the theorem statement was independently checked by Claude Code Opus 5.5. The equal-degree reductions in Section~\ref{sec:equal-degree}\ifrmext{} and the Reed--Muller extension in Section~\ref{sec:rm} and Appendix~\ref{app:rm-proof}\fi{} are not covered by the Lean formalization. The authors take full responsibility for the contents of this paper.

\clearpage
\begingroup
\raggedright
\bibliographystyle{alphaurl}
\bibliography{references}
\endgroup

\ifrmext
\appendix
\section{Proof of the Reed--Muller extension}\label{app:rm-proof}
The proof of Theorem~\ref{thm:rm} follows Section~\ref{sec:main-proof},
and only the steps that change are described here. A pair
$(\boldsymbol\alpha,P)$ is \emph{exceptional} if $\deg P\le d$,
$|S_{\boldsymbol\alpha}(P)|\ge A$, and $P$ has no admissible witnesses
with agreement loss at most $n/q_0$. As in Section~\ref{sec:main-proof},
we divide the coefficient vectors $\boldsymbol\alpha\in\F_q^{k+1}$ for
which some $(\boldsymbol\alpha,P)$ is exceptional into the sets
$\mathcal E$, $\mathcal P$, and $\mathcal N$ of excluded, persistent, and
nonpersistent ones, and bound each set separately.

\paragraph{Reduction to univariate polynomials.}
Following~\cite{PW04}, we map the Reed--Muller code into a Reed--Solomon
code with the same degree fraction by identifying the grid with the
field $\F_n$ through an $\F_{q_0}$-linear bijection
$\psi=(\psi_1,\ldots,\psi_k)$. Its components
$\psi_i(X)=\operatorname{Tr}_{\F_n/\F_{q_0}}(\gamma_iX)$, for a basis
$\gamma_1,\ldots,\gamma_k$ of $\F_n$ over $\F_{q_0}$, are polynomials
of degree $n/q_0$. Substituting $\psi_1,\ldots,\psi_k$ for
$X_1,\ldots,X_k$ maps each polynomial $P$ of degree at most $d$ to a
univariate polynomial $\widetilde P$ of degree at most $\rho n$, and
distinct $P$ give distinct $\widetilde P$: if
$\widetilde P_1=\widetilde P_2$, then $P_1-P_2$ vanishes at all $n$ grid
points, and since its total degree is less than $q_0$, $P_1=P_2$ by the
Schwartz--Zippel lemma~\cite{Sch80,Zip79}. With the
received words $f\circ\psi$ and $g\circ\psi$ on $\F_n$, each candidate
$P$ corresponds to the Reed--Solomon candidate $\widetilde P$, with the
same number of agreements, for the multiplier
$\widetilde L_{\boldsymbol\alpha}=\alpha_0+\sum_i\alpha_i\psi_i$, which has
degree $n/q_0$. The interpolation points therefore lie in $\F_n$, while
the received words take values in $\F_q$, so we work over the smallest
field $\F'$ containing $\F_q$ and $\F_n$. Up to the
agreement-incidence argument, Section~\ref{sec:main-proof} uses no
property of its base field other than being finite, so the factorization
of $Q$, the normalization by $e_i$-th powers, the bound on $\mathcal E$,
and simple-root lifting hold over $\F'$, with
$K\coloneq\overline{\F'(T_0)}$. Claim~\ref{lem:agreement-incidence} is
the exception, treated below.

\paragraph{Interpolation.}
The interpolation follows Lemma~\ref{lem:roots} over $\F'$, with the
$(1,n/q_0,\rho n)$-weighted degree. With $m$ and $D$ from Section~\ref{sec:interp}, the $Z$-degree
is bounded by
\[
 M\coloneq\min\left\{D,\floor{\frac{mA-(n/q_0)D-1}{\rho n}}\right\}.
\]
Since $\deg\widetilde L_{\boldsymbol\alpha}\le n/q_0$, the condition
$D\le mA-U$ in the proof of Lemma~\ref{lem:roots}, where
$U\coloneq\ceil{(m+1/2)\sqrt\rho\,n}$, becomes $(n/q_0)D\le mA-U$. In
that proof, $mA-U\ge m\eta n/2-1$; hence this condition holds when
$q_0\eta^2\ge4/\rho$, because $D+1\le2m^2/(3\rho)$ and $m<3/\eta$. The
result is a nonzero polynomial $Q$ such that
$Q(X,\widetilde L_{\boldsymbol\alpha}(X),\widetilde P(X))=0$ whenever
$\deg P\le d$ and $|S_{\boldsymbol\alpha}(P)|\ge A$.

\paragraph{Excluded coefficient vectors.}
We factor $Q$ as in
Section~\ref{sec:main-proof} and define $\mathcal E$
by~\eqref{eq:excluded-coefficients}, with
$\widetilde L_{\boldsymbol\alpha}$ instead of $\alpha+\beta X$ and with
the $e_i$-th root of $\widetilde L_{\boldsymbol\alpha}(X^{e_i})$ instead
of $\alpha^{1/e_i}+\beta^{1/e_i}X$. Then $|\mathcal E|\le D^2$ by the
proof of Lemma~\ref{lem:excluded}, which only uses that distinct
coefficient vectors give distinct polynomials
$\widetilde L_{\boldsymbol\alpha}$; this follows from the argument above
with $P=L_{\boldsymbol\alpha}$.

\paragraph{Persistent coefficient vectors.}
As in Section~\ref{sec:persistent}, we first bound the number of
exceptional coefficient vectors for one fixed rational pair.

\begin{lemma}\label{lem:rm-rational-pair}
Let $F,G\in\F_q(\mathbf X)$. The number of coefficient vectors
$\boldsymbol\alpha$ for which $P\coloneq F+L_{\boldsymbol\alpha}G$ is a
polynomial and $(\boldsymbol\alpha,P)$ is exceptional is at most
$\binom n{k+1}$ if $F,G$ are polynomials with $\deg F\le d$ and
$\deg G\le d-1$, and at most $q$ otherwise.
\end{lemma}
\begin{proof}
First suppose that $F,G$ are polynomials with these degree bounds, and
let $I'$ be their full correlated agreement set with $f,g$. At an accidental
agreement $\mathbf z\notin I'$, we have $g(\mathbf z)\ne G(\mathbf z)$ as
in~\eqref{eq:extra}, so the agreement imposes the linear condition
\[
 L_{\boldsymbol\alpha}(\mathbf z)
 =\frac{F(\mathbf z)-f(\mathbf z)}{g(\mathbf z)-G(\mathbf z)}
\]
on $\boldsymbol\alpha$. Since $F$ and $G$ are fixed, the right-hand side
does not depend on $\boldsymbol\alpha$. If
$(\boldsymbol\alpha,P)$ is exceptional, then $F,G$ are not admissible
witnesses, so $P$ has more than $n/q_0$ accidental agreements.
By~\eqref{eq:rm-hyperplane}, these include $k+1$ affinely independent
points, and the conditions at these points determine
$\boldsymbol\alpha$. Choosing such $k+1$ points for each exceptional
coefficient vector therefore assigns distinct sets to distinct vectors,
and there are at most $\binom n{k+1}$ sets.

Otherwise, the coefficient vectors for which $F+L_{\boldsymbol\alpha}G$
is a polynomial of degree at most $d$ form an affine subspace of
$\F_q^{k+1}$, if nonempty, and we show that it has dimension at most
one. In Lemma~\ref{lem:rational-pair}, it sufficed that the subspace is
proper, but a proper subspace of $\F_q^{k+1}$ can contain $q^k$
coefficient vectors. Note that $G$ is not a polynomial of degree at most
$d-1$, since then $F=P-L_{\boldsymbol\alpha}G$ would be a polynomial of
degree at most $d$, which is the first case. Now let
$\boldsymbol\alpha\ne\boldsymbol\alpha'$ lie in the subspace, so that
$L_{\boldsymbol\alpha-\boldsymbol\alpha'}G$ is a polynomial of degree at
most $d$. If $G$ is not a polynomial, its denominator in lowest terms
divides the nonzero polynomial $L_{\boldsymbol\alpha-\boldsymbol\alpha'}$
of degree at most one, so the denominator is $L_{\boldsymbol\sigma}$ for
some nonzero $\boldsymbol\sigma$ determined by $G$, and
$\boldsymbol\alpha-\boldsymbol\alpha'$ is a multiple of
$\boldsymbol\sigma$. If $G$ is a polynomial of degree at least $d$, then
$L_{\boldsymbol\alpha-\boldsymbol\alpha'}$ must be constant, so
$\boldsymbol\alpha-\boldsymbol\alpha'$ is a multiple of
$(1,0,\ldots,0)$. In both cases, all differences between elements of the
subspace are multiples of one vector, so the subspace has dimension at
most one.
\end{proof}

Persistence is defined as in Section~\ref{sec:main-proof}, with rational
pairs $F,G$ such that $G=H/L_{\boldsymbol\sigma}$ for a polynomial $H$ of
degree at most $d$ and a nonzero $\boldsymbol\sigma\in\F_q^{k+1}$. The
pair defined in the proof of Claim~\ref{lem:agreement-incidence}
from two solutions $(\boldsymbol\alpha,P)$ and $(\boldsymbol\alpha',P')$
is of this form, with $H=P'-P$ and
$\boldsymbol\sigma=\boldsymbol\alpha'-\boldsymbol\alpha$. There, the pair obtained from $c$ agreement hyperplanes
agrees with $f,g$ on the $c$ corresponding coordinates, except at the
zeros of $L_{\boldsymbol\sigma}$. If $L_{\boldsymbol\sigma}$ is not
constant, its zeros form a hyperplane of $\F_q^k$, so
by~\eqref{eq:rm-hyperplane}, $L_{\boldsymbol\sigma}$ has at most $n/q_0$
zeros on the grid, while the denominator $L$ in
Section~\ref{sec:main-proof} has at most one zero. We therefore set
$B\coloneq\ceil{(\sqrt\rho+\eta/2)n}-n/q_0$. For two
distinct pairs $(F_j,G_j)$ of this form, with
$G_j=H_j/L_{\boldsymbol\sigma_j}$, explaining exceptional
$(\boldsymbol\alpha_j,P_j)$ for $j=0,1$, the intersection of their
correlated agreement sets lies in the zero set of a nonzero polynomial of
degree at most $d+1$, as in the proof of Lemma~\ref{lem:persistent}. By
the Schwartz--Zippel lemma, this intersection therefore contains at most
$(d+1)n/q_0$ coordinates. Since
$B>(d+1)n/q_0$, as shown below, Lemma~\ref{lem:johnson} with $a=B/n$
and $b=(d+1)/q_0$ bounds the number of such pairs by
$\ell\coloneq(a-b)/(a^2-b)$, and by Lemma~\ref{lem:rm-rational-pair},
$|\mathcal P|\le\ell\max\{q,\binom n{k+1}\}$.

\paragraph{Nonpersistent coefficient vectors.}
Simple-root lifting applies to $\widetilde L_{\boldsymbol\alpha}$ as to
the line $\alpha+\beta X$. For a solution $(\boldsymbol\alpha,P)$, set
$Y\coloneq\widetilde L_{\boldsymbol\alpha}(T_0^{e_i})^{1/e_i}$,
$V_j\coloneq\alpha_j^{1/e_i}$, and $Z\coloneq\widetilde P(T_0^{e_i})$. At
$X=T_0+W$, the $e_i$-th root of $\widetilde L_{\boldsymbol\alpha}(X^{e_i})$
equals $Y+\sum_jV_j\theta_j(W)$ for fixed polynomials
$\theta_1,\ldots,\theta_k\in K[W]$ with $\theta_j(0)=0$.
The proof of Lemma~\ref{lem:hensel-lifting} also applies to this
polynomial. In the proof, the coefficients of $Y+V(T_0+W)$ in $W$
are linear in $(Y,V)$, and here the coefficients of
$Y+\sum_jV_j\theta_j(W)$ are linear in $(Y,V_1,\ldots,V_k)$. The
parameters $(Y,V_1,\ldots,V_k,Z)$ lie on the hypersurface
$R_i(T_0,Y,Z)=0$ in $K^{k+2}$, of dimension $k+1$. The reconstruction in
Claim~\ref{lem:envelope}, with $\rho n$ instead of $d$, therefore
produces algebraic sets of dimension at most $k+1$ and degree at most
$h_i(2e_i\rho nh_i)^{k+1}$.

Claim~\ref{lem:agreement-incidence} cannot be applied to these sets
directly. Its proof needs agreement hyperplanes defined over $\F_q$, so
that the rational pair it constructs is defined over $\F_q$, but in the
coefficient space of $(\boldsymbol\alpha,\widetilde P)$, the agreement
condition at a point $x\in\F_n$ involves the powers $x^j\in\F_n$. The
sets are therefore moved to the coefficient space
$K^{k+1}\times K^{\binom{d+k}{k}}$ of $(\boldsymbol\alpha,P)$, where the
agreement hyperplanes
$P(\mathbf z)=f(\mathbf z)+L_{\boldsymbol\alpha}(\mathbf z)g(\mathbf z)$
are defined over $\F_q$. To do so, we intersect them with the image of
the injective linear map
$(\boldsymbol\alpha,P)\mapsto(\boldsymbol\alpha,\widetilde P)$, which also
discards reconstructed polynomials not of the form $\widetilde P$, and
pull the intersection back along this map. By
Lemma~\ref{lem:hyperplane-section}, applied to each hyperplane defining
the image, the intersection does not increase the dimension or the
degree, and the pullback preserves both. The claim then applies with two
changes to its proof: $\boldsymbol\alpha\ne\boldsymbol\alpha'$ by the
Schwartz--Zippel lemma, and the two cases are $c>B+n/q_0-1$ and
$c\le B+n/q_0-1$, so that in the first case the pair agrees with $f,g$
on at least $c-n/q_0\ge B$ coordinates outside the zeros of
$L_{\boldsymbol\sigma}$. Summing over the factors as in the proof of
Lemma~\ref{lem:nonpersistent}, with $v\coloneq n/(A-B-n/q_0+1)$, we
obtain $|\mathcal N|\le(2\rho nMv)^{k+1}D^{k+2}$.

\paragraph{The bound.}
As in the proof of Lemma~\ref{lem:counting}, with $\rho n$ instead of
$d$, the parameters satisfy $D\le9/(2\rho\eta^2)$, $\rho nM\le3n/\eta$,
$v\le2/\eta$, and $\ell\le4/\eta$. For $v$, note that
$A-B-n/q_0+1=A-\ceil{(\sqrt\rho+\eta/2)n}+1>\eta n/2$. For $\ell$, the
condition $q_0\eta^2\ge4/\rho$, which also forces $q_0>64$, implies
$a-\sqrt b\ge\eta/4$ by the same computation; in
particular, $B>(d+1)n/q_0$. Every coefficient vector we count lies in
$\mathcal E\cup\mathcal P\cup\mathcal N$, so their number is at most
\[
 D^2+\ell\max\{q,\tbinom n{k+1}\}+(2\rho nMv)^{k+1}D^{k+2}
 \le\frac{4q}{\eta}+\frac{2^{6k+8}n^{k+1}}{\rho^{k+2}\eta^{4k+6}}.
\]
Dividing by $q^{k+1}$ proves Theorem~\ref{thm:rm}.
\fi

\end{document}